***Plasmodium falciparum* Hop:**

***Detailed analysis on complex formation with Hsp70 and Hsp90***

Rowan HATHERLEY, Crystal-Leigh CLITHEROE, Ngonidzashe FAYA and Özlem TASTAN BISHOP*

Research Unit in Bioinformatics (RUBi), Department of Biochemistry and Microbiology, Rhodes University, 6140, Grahamstown, South Africa.

*Correspondence to Özlem TASTAN BISHOP; Research Unit in Bioinformatics (RUBi), Department of Biochemistry and Microbiology, Rhodes University, 6140, Grahamstown, South Africa; phone: +27 46 603 8072; e-mail: O.TastanBishop@ru.ac.za

## Abstract

The heat shock organizing protein (Hop) is important in modulating the activity and co-interaction of two chaperones: heat shock protein 70 and 90 (Hsp70 and Hsp90). Recent research suggested that *Plasmodium falciparum* Hop (PfHop), PfHsp70 and PfHsp90 form a complex in the trophozoite infective stage. However, there has been little computational research on the malarial Hop protein in complex with other malarial Hsps. Using *in silico* characterization of the protein, this work showed that individual domains of Hop are evolving at different rates within the protein. Differences between human Hop (HsHop) and PfHop were identified by motif analysis. Homology modeling of PfHop and HsHop in complex with their own cytosolic Hsp90 and Hsp70 C-terminal peptide partners indicated excellent conservation of the Hop concave TPR sites bound to the C-terminal motifs of partner proteins. Further, we analyzed additional binding sites between Hop and Hsp90, and showed, for the first time, that they are distinctly less conserved between human and malaria parasite. These sites are located on the convex surface of Hop TPR2, and involved in interactions with the Hsp90 middle domain. Since the convex sites are less conserved than the concave sites, it makes their potential for malarial inhibitor design extremely attractive (as opposed to the concave sites which have been the focus of previous efforts).

## Introduction

Heat shock proteins (Hsps) are an important class of molecular chaperones. Hsp70 and Hsp90 are potential drug targets for inhibitor design to control the increasingly drug resistant malaria parasite *Plasmodium falciparum* (*P. falciparum*) [1–3]. A less well studied co-chaperone, the Hsp70/Hsp90 organizing protein, Hop, plays an important role in modulating the activity and co-interaction of these two chaperones [4,5].

Hop contains three tetratricopeptide repeat (TPR) domains, each having three TPR motifs and two DP domains; DP1, between TPR1 and TPR2A; and DP2, between TPR2B and the C-terminal end of Hop [4]. DP2 comprises five helices, forming an elongated V-shape. DP1 consists of roughly the same five helices, but with a short additional helix near the N-terminus, and is more globular than DP2 [4]. To date, there is no structure for the whole Hop protein, however there have been studies to discern Hsp70:Hop and Hsp90:Hop complexes [4–8].

The surface of each TPR domain has concave and convex regions [9]. The best understood aspects of Hop interactions are those that occur at the concave regions: binding to specific C-terminal motifs in Hsp70 and Hsp90. TPR1 domain binds the C-terminal Hsp70 peptide motif, EEVD [6–8] and Hsp104 [10]; TPR2B domain also binds the EEVD residues of Hsp70 [4–8]; and TPR2A domain binds the MEEVD C-terminal residues of Hsp90 [7,8,11,12]. However, binding and interaction energy studies indicated that the affinity of the concave interface of TPR2A for Hsp90 C-terminal peptide does not explain the full affinity of Hop for Hsp90 [13]. Recent work on *Saccharomyces cerevisiae* (*S. cerevisiae*) showed that the convex surfaces of TPR2A and TPR2B interact with the surface of the Hsp90 middle (M) domain, while the Hsp90 C-terminal (C) domain interacts with the peptide-binding groove of TPR2A [4]. Interestingly, it is thought that interaction with convex regions of TPR2 forces Hsp90 into a conformation that prevents its ATPase activity [4,5]. Protein structures for the Hop TPR domains and the respective C-terminal motifs of Hsp70 and Hsp90 in complex are known for only *Homo sapiens* [7] and *S. cerevisiae* [4]. Recently, it was suggested that Hop forms a complex with Hsp90 and Hsp70 in the *P. falciparum* trophozoite (within the infected host erythrocyte), and is overexpressed in this infective stage [14]. This shows the importance of the protein for analysis as a potential drug target.

This work performed detailed *in silico* characterization and comparison of human and *P. falciparum* Hop proteins (HsHop and PfHop respectively), and relevant interactions with other Hsps, to determine prospective inhibitor sites. The research was divided into two parts. Firstly, Hop was analyzed on a large scale to assess its variability across species. Secondly, 3D models of HsHop and PfHop domains and their complexes with cytosolic variants of Hsp70 and Hsp90 proteins were calculated and studied in terms of structure and interaction sites. The results, for the first time, indicated that there are variable sites in the convex regions of Hop involved in complex formation with Hsp90. As convex sites are less conserved than concave regions, they are attractive for malarial inhibitor design.

## Methods and Materials

### *Sequence acquisition*

PfHop sequence (PF3D7_1434300) was retrieved from PlasmoDBv9.3 [15], and homolog Hop proteins were searched by NCBI-BLAST with an E value cut-off of > $10^{-50}$. The rigid E value was selected to filter out non-Hop homologs with TPR regions, as this is a common motif in a number of other proteins as well. 88 sequences were retrieved (S-Data 1). Prokaryote sequences were not selected, as the focus of this study was eukaryotic organisms. Further, PfHsp90 (PF3D7_0708400), PfHsp70-1 (PF3D7_0818900), HsHsp90 alpha (NP_005339.3) and HsHsp90 beta (NP_031381.2) and HsHsp70-1A (NP_005337.2) were obtained for homology modeling.

### *Sequence and motif analysis*

Multiple sequence alignment (MSA) was done by MAFFT's E-INSi's protocol [16]. All versus all pairwise sequence identity scores were calculated for each domain, and the distribution was displayed using MATLAB scripts.

The protein sequences were submitted to the MEME web-server [17] to search for conserved motifs. Motifs with length range 2 – 150 amino acids were searched as this is the approximate length of the longest domains in Hop (TPR domains).

### *Homology modeling and validation*

Initial modeling was done to investigate the interaction interface between the convex regions of the Hop TPR2A and TPR2B domians (HopTPR2AB) with Hsp90 M domain, based on *S. cerevisiae* work by Schmid *et al* [4]. The template used to model this interaction was the structure of ScHopTPR2AB domain complexed with ScHsp90 M and C domains (ScHopTPR2AB-ScHsp90MC). The coordinate file was kindly supplied by Schmid *et al.* [4] and was calculated via docking an ScHopTPR2 fragment (PDB ID: 3UQ3) to a low-resolution spin-labeled model ScHsp90 M and C domain fragment. In this structure, Schmid *et al.* exchanged Hsp90 residues S-184, S-411 and S-422 with CXM residues (Cysteine with an additional proxyl group). For modeling, two templates were prepared: Firstly, the CYS-template, where CXM residues were converted to Cysteine residues; and secondly, a SER-template, where each of these residues in the CYS-template was converted to Serine, using MODELLER's mutate-function. Multi-chain homology modeling employed MODELLER 9.12 [18] according to [19]. For each protein (or protein complex) 100 models were built. The best 5 models from each set were chosen based on DOPE Z-score, and submitted to MetaMQAPII for further evaluation, yielding the top model for each set based on the GDT_TS score and the local structural quality at the predicted interaction interface (based on the ScHopTPR2AB-ScHsp90MC template).

Interactions between concave regions of Hop TPR domains with the C-terminal peptides of Hsp70 and Hsp90 were also modeled, using templates 1ELW, 3UQ3 and 3UPV. 1ELW was used to model HopTPR1 with residues GPTIEEVD from the C-terminus of Hsp70 (HopTPR1-Hsp70$_{GPTIEEVD}$). 3UQ3 was used to model HopTPR2AB with both MEEVD from the Hsp90 C-terminus, as well as EVD from the Hsp70 C-terminus (HopTPR2AB-Hsp70$_{EVD}$-Hsp90$_{MEEVD}$). In this complex, HopTPR2A interacts with Hsp90$_{MEEVD}$ and HopTPR2B interacts with Hsp70$_{EVD}$. 3UPV was used to model HopTPR2B with C-terminal residues PTVEEVD from Hsp70 (HopTPR2B-Hsp70$_{PTVEEVD}$). Modeling was performed as with the complexes used to study the Hop convex interactions.

***Protein-protein docking***

Since the ScHopTPR2AB-ScHsp90MC template, used to model the complexes, was determined by docking [4], each complex (including the template) was resubmitted to HADDOCK [20] for docking to confirm the orientation of the interaction interface. Active residues set when docking were those that formed interactions in the ScHopTPR2AB-ScHsp90MC template, with passive residues being assigned by the server.

***Identification of important residues in complexes***

Complexes were submitted to the Protein Interaction Calculator (PIC) web-server [21], andPython scripts were written to extract the conserved interacting residues (defined as those that interacted in more than half of the complexes observed). Then for each conserved residue, the script used the PIC results and recorded which residues it interacted with. Using the same conservation level cut-off, conserved interaction networks for the different groupings were created.

Each HopTPR2AB-Hsp90MC complex was also submitted to the Robetta Alanine Scanning web-server [22]. Python scripts recorded the conserved “hot spot” residues for each grouping. The binding hot spot cut off value used was a $\Delta\Delta$Gbind $\geq$ 1.0 kcal.mol$^{-1}$ [23].

## Results and Discussion

### ***Individual domains evolve at different rates within Hop:***

Sequence identities for each possible pair of 88 sequences for each of the Hop domains are shown in Fig. 1, both as matrices and histograms. In the histograms, the pairwise alignment score of each sequence against itself was not represented. Overall, the results showed that Hop domains are highly diverse and individual domains evolve at different rates within the protein. DP1 was the least conserved region with most pair identities between 0.2 and 0.4, which was followed by DP2 (0.3 – 0.5). TPR1, TPR2A displayed most pair identities between 0.4 and 0.6, while the most conserved domain was TPR2B, with most pair identities above 0.5. The matrices demonstrated some phylogenetic clustering; e.g., the well conserved block (red and yellow) in the middle of each matrix represented the Kingdom Animalia, and the extremely well conserved block (red) representing the mammals.

### ***Motif analysis reflected domain separation:***

Motif analysis was applied to the full-length Hop sequence dataset, and up to 30 motifs were searched. TPR domains are relatively well described in the literature [4–8]. MEME results reflected this separation clearly (S-Data 2A). Each domain was represented either by one long motif or by a combination of smaller motifs. The conservation of the motifs throughout sequences is presented in Fig. 2. Motifs 1, 2, 3, 5 and 6 were present in all sequences. The most conserved region of Hop, TPR2B was represented by a single motif, motif 1. Motif 2 was part of the TPR1 domain. Interestingly, DP2 was also represented by a single motif, motif 3. On the other hand, TPR2A and linker region were presented by more than one motif (4, 5, 6), but conserved in all organisms. The motif conservation of DP2 may reflect the functional importance of this domain, as previously demonstrated [4,24–26]. Conversely, the DP1 domain and linker region connecting DP1 appears to be the least well conserved region in the protein. DP1 was represented with different motifs (e.g. 7, 9, 10, 12, 16 and 18) in different organisms, which explains why it was the least conserved in sequence identity calculations. In most species, the DP1 domain possesses recognizable “DP/NP” repeats (as are found in the DP2 domain) and the long linker contains proline repeats. However this is not so in the apicomplexan taxa (particularly in the *Plasmodium* genus), and DP1 has yet to be characterized experimentally in PfHop. All *Plasmodium sp.* sequences and a few protozoan Hop sequences uniquely lacked DP1 motifs 7 and 10, which were present in the other sequences, including human Hop. Motif 7 is located in the DP1 domain while motif 10 is shared by DP1 and linker regions. In addition, motif 11, 13, 16 and 20 were present in the HsHop sequence while absent in the *Plasmodium sp*. Motif 12, 14, 23, and 30 were observed to be present in *Plasmodium sp*. but absent in HsHop.

Even though motif analysis of the TPR2 domain shows this region is highly conserved, at a residue level there are striking differences between human and *P. falciparum* Hop sequences, and especially in the convex regions where the protein is interacting with Hsp90, as discussed later in this article.

***Convex regions of TPR2 are potential target sites for inhibitor design:***

The concave and convex regions of TPR2 (Fig. 3) and their interactions with Hsp70 and Hsp90 are of great interest. As part of the analysis, Hsp90MC-HopTPR2AB complexes were generated by modeling and HADDOCK docking for PfHsp90 (PfHsp90MC-PfHopTPR2AB), HsHsp90-alpha (HsHsp90αMC-HsHopTPR2AB) and HsHsp90-beta (HsHsp90βMC-HsHopTPR2AB), with their respective potential Hop counterparts (see S-Data 3 for model quality and docking quality assessments). The orientation of the template complex used for modeling was reproduced when the monomers of each complex were re-docked, with HADDOCK scores comparable to the re-docked templates (S-Data 3). Interactions calculated between the complexes are displayed in Fig. 4. The residues are numbered to allow comparison of the proteins. For actual residue numbers, refer to S-Data 4A.

Although both the M and C domains of Hsp90 were present in these models, the interaction interface was limited to Hsp90 M domain only. The common interactions between HopTPR2AB and the Hsp90 M domains across the different sets of complexes are tabulated in S-Data 4B and summarized in Fig. 4. When comparing the human models to the yeast template, there was a great deal of overlap, with 18 interactions common to these three sets of complexes, with 3 specific to the human complexes and 7 that occurred in the yeast template complexes only. When comparing the human complexes to those of *P. falciparum*, the number of common interactions drops to 12, with 9 interactions found in both human complexes, but not in the *P. falciparum* complexes and 11 found only in the *P. falciparum* complexes. It was found that the human complexes seemed to be stabilized by a number of hydrophobic interactions between residue W28 of HsHsp90 and M156, Y180 and L188 of HsHop. All three residues differed in PfHop and only L180 formed hydrophobic interactions with W28 of PfHsp90. Another interesting difference is the ionic interaction observed between HsHsp90 E140 and HsHop K123 (present as E123 in PfHop). In the *P. falciparum* complexes this ionic interaction was shifted, causing K143 of PfHsp90 to form ionic interactions with both E119 and E123 of PfHop. PfHsp90 K177 formed hydrogen bonds with PfHop R109. These residues differed in both HsHsp90 and HsHop, so there was no observed hydrogen bonding between these residues. Similarly, PfHsp90 D180 interacted ionically with PfHop K116, but this was not seen in the human models due to HsHop having residue Q116 in this position. PfHsp90 D187 formed both ionic interactions and hydrogen bonds with PfHop R112 (L112 in the HsHop models). Finally, ionic interactions between D190 of PfHsp90 and R112 of PfHop were not observed in their human counterparts, as these Hsp90 residues were present as T190 in HsHsp90-alpha and S190 in HsHsp90-beta. This final interaction pair may be significant. The results of computing alanine scanning (S-Data 4C), indicated that R191 of Hsp90 was the only hot spot residue consistently identified across all complexes, even though it was not directly involved in any interactions. Hsp90 residue W28 has been shown to be important to Hsp90-Hop interactions in yeast [27], and was identified as a hot spot residue in yeast and human complexes, but not in *P. falciparum*. In *P. falciparum* complexes, there appeared to be fewer hydrophobic interactions with this residue (Fig. 4). For the Hop residues, R176 was the only hot spot residue identified in all complexes. This formed a cation-π interaction with W28. Residue N108 of PfHop (T108 in HsHop) was found to be a hot spot residue exclusively in *P. falciparum*. This residue forms hydrogen bonds with E194 of PfHsp90, which was also identified as a hot spot residue in *P. falciparum* and not human Hsp90s. PfHop residue R112 (L112 in HsHop) was identified as a hot spot residue in PfHop exclusively. This formed interactions with E187 and D190 of PfHsp90, neither of which was found in human complexes. Both residues N108 and R112 of PfHop were found to be hot spot residues exclusively in *P. falciparum* complexes and both these residues differ in HsHop. These may be good candidates for further investigation as target sites for potential inhibitor design.

### ***Motifs 1, 6 and 20 form the convex interaction interface***

To further investigate evolutionary conservation, especially between human and *P. falciparum*, motifs representing the TPR2 domains (in both organisms) were mapped to the PfHsp90MC-PfHopTPR2AB interaction interface and analyzed (S-Data 2C). Motif 1 (TPR2B domain) is 113 residues long (S-Data 2B), covering positions 127 – 240 in our models. HsHop and PfHop share 57% sequence identity along this region and these identical residues were mapped to the complex structure (S-Data 2C). Although the interface between Hop and Hsp90 has a large number of identical residues, some differences were observed. In spite of its large size, only four residues from motif 1 of PfHop (residues 156, 157, 176 and 180) interacted with PfHsp90. These residues formed interactions in HsHop, in addition to five other flanking residues (Fig. 4). Only two of the four common residues were found to be identical in both HsHop and PfHop. These residues formed similar interactions with Hsp90 (Fig 4). Interestingly, analysis of motif 6 (part of TPR2A) gave the opposite result. This motif is 29 residues long (S-Data 2B) and only 9 residues are identical (31%) between human and *P. falciparum*. The identical residues are mostly located away from the interface. In fact, S-Data 2C shows a portion of the interface in which there are no conserved residues between human and *P. falciparum*. Of the interacting residues described above, eight corresponded to this motif in PfHop and seven to HsHop (positions 92 – 120 in our models). Of these residues, five interacted in the same position in both PfHop and HsHop (residues 108, 113, 116, 119 and 120; Fig. 4) only one of which, E119, was identical in both PfHop and HsHop (S-Data 4A). Finally, motif 20 was only identified in human Hop and covered positions 121 – 126 in our models. As discussed above, residue 123 interacts in both PfHop and HsHop, but the difference in residue type, seems to have shifted this interaction down in PfHop, so that residues 119 and 123 interact with residue 143 of PfHsp90, rather than residues 139, 140 and 143, as seen in the human counterpart (Fig. 4). Overall, it appears that Hop motifs 1, 6 and 20 are adjacent motifs that combine to form the convex interaction interface with Hsp90.

***Interactions in the concave regions of Hop TPR domains are conserved***

As much work has focused on interactions in the concave regions of TPR1 and TPR2B (with Hsp70 C-terminal peptide) and TPR2A (with Hsp90 C-terminal peptide), these were also analyzed. The results are summarized in S-Data 5A. It was found that in each TPR domain, there were many interactions common to both human and *P. falciparum* and relatively few that were specific to only human or *P. falciparum*. The closest exception to this was the TPR1-Hsp70$_{GPTIEEVD}$ interaction, which involved more hydrophobic interactions in the human set; however, even with these, there were still more interactions common to both human and *P. falciparum*. Even when comparing results from different TPR domains, there were a number of interactions conserved in all three domains (TPR1, TPR2A and TPR2B) (S-Data 5B and 5C). The results indicate that the way Hop TPR domains interact with Hsp70/Hsp90 C-terminal residues is somewhat conserved even across different TPR domains.

To date, attention in the literature has been mainly focused on these concave sites shown in Fig. 3. Here, for the first time, interactions between the convex regions of HopTPR2AB and the Hsp90 M domains (Fig. 3) were identified for both human and parasite, and were shown to be less conserved, making these regions potential sites for inhibitor design (Fig. 4, S-Data 4).

## Conclusion

Hop plays an important role in modulating the activity and co-interaction of two chaperones, Hsp90 and Hsp70, yet little is known about it. In this study a number of novel results were obtained. *In silico* studies showed that Hop is highly conserved in overall structure, and that the individual domains of Hop evolve at different rates within the protein. Overall, the most well conserved region of Hop was the TPR2B domain, both at sequence and motif levels. Other than the TPR2B domain, DP2 was the only domain represented by a single motif, reflecting the known functional importance of this domain. Conversely, the DP1 domain and linker region connecting DP1 and TPR2A domain, together, formed the least well conserved region. MSA indicated that this region may form a structurally and functionally distinct part that is unique to apicomplexan taxa, and has yet to be characterized experimentally in PfHop. Comparative interaction studies in both Pf and HsHop suggested that the concave site TPR residues involved in interaction with C-terminal partner peptides are far more evolutionarily stable than those at the convex sites on TPR2 that interact with Hsp90 M domain. Previous work on Hop has investigated the concave sites for drug targeting; but the convex Hsp90 interacting sites are less conserved than the concave ones, thus making them particularly attractive for malarial inhibitor design.

## Acknowledgements

R.H. thanks the National Research Foundation (NRF; 79765), South Africa, for financial support. Opinions expressed and conclusions arrived at are those of the author and are not necessarily to be attributed to the NRF. C-L.C. and N.F. thank Rhodes University for financial support.

## Figure Legends:

**Fig. 1: Distribution of pairwise identity scores per each Hop domain.** The top row displays scores as a matrix (identity scores for every sequence versus every sequence represented as a fraction as per the scale), bottom row as histograms, where the x-axis represents sequence identity as a fraction and the y-axis represents number of sequence pairs. A) TPR1, B) DP1, C) TPR2A, D) TPR2B and E) DP2.

**Fig. 2: Motif conservation presented as a heat map.** Conservation increases from blue to red while white color represents the absence of a motif.

**Fig. 3: Cartoon representation of convex and concave interaction sites of the TPR2 domain.** (A) Hsp90 M and C domains are shown in red (cartoon representation); HopTPR2 domain is shown in green (cartoon representation); Hsp90 (red) and Hsp70 (blue) C-terminal peptides are presented as spheres. (B) The TPR2 domain is rotated 90° in order to show the concave interaction sites more clearly.

**Fig. 4: Consensus interaction network diagram of human, *P. falciparum* and yeast HopTPR2AB-Hsp90MC interfaces studied.** Hsp90 residues are shown as blue ovals, mapped to their positions on the cartoon structure segments, rendered in PyMOL. Hop residues are displayed as red squares. Residue interaction types are displayed as follows: solid line) ionic or cation-π interaction; dotted line: hydrogen bonding; faded triangle: hydrophobic interactions. Residues without interaction lines displayed non-specific interactions. Each panel represents the following complexes: A) HsHsp90αMC-HsHopTPR2AB; B) HsHsp90βMC-HsHopTPR2AB; C) PfHsp90MC-PfHopTPR2AB; D) Schmid *et al*. template (ScHopTPR2AB-ScHsp90MC). Residue numbers are shown based on their positions in the models created, to make them comparable. For actual residue numbers, refer to S-Data 4A.

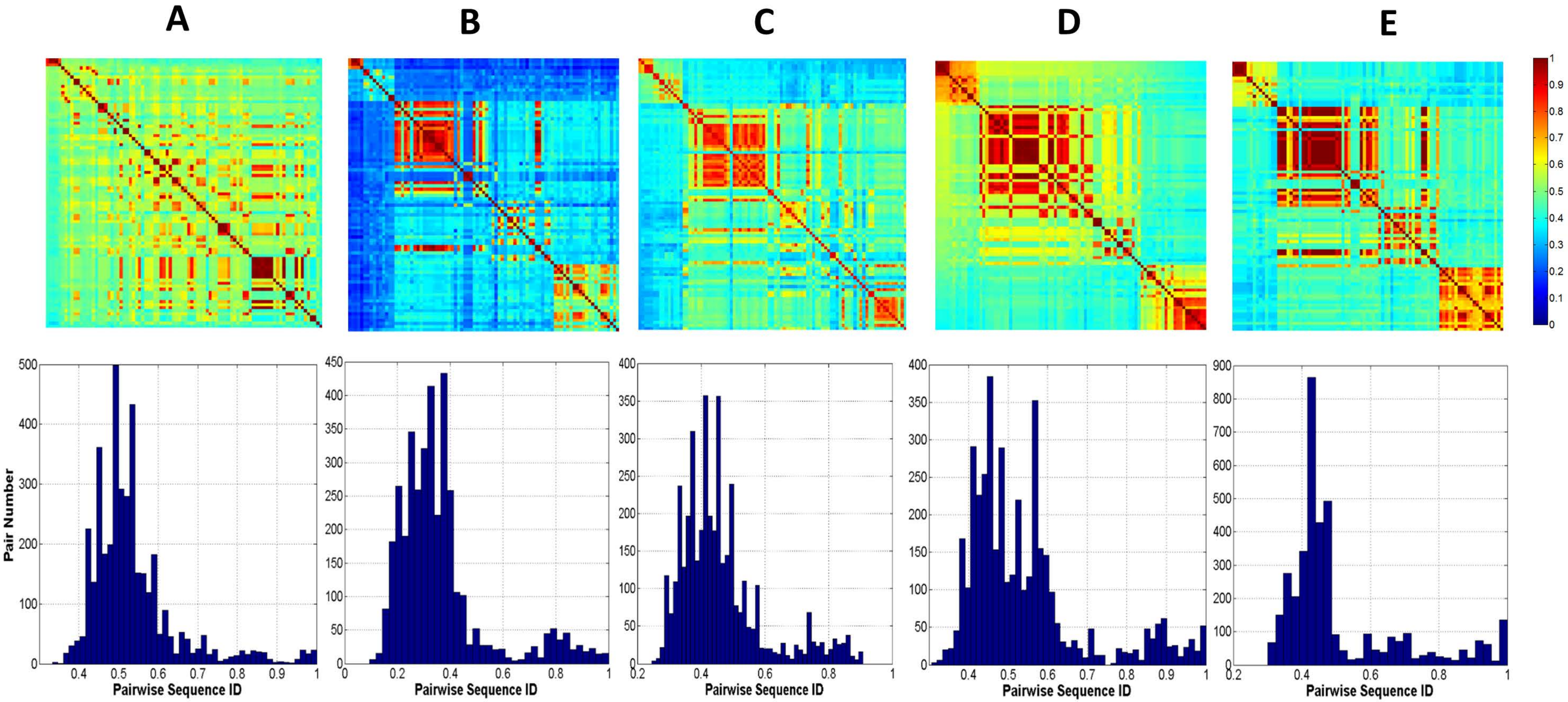

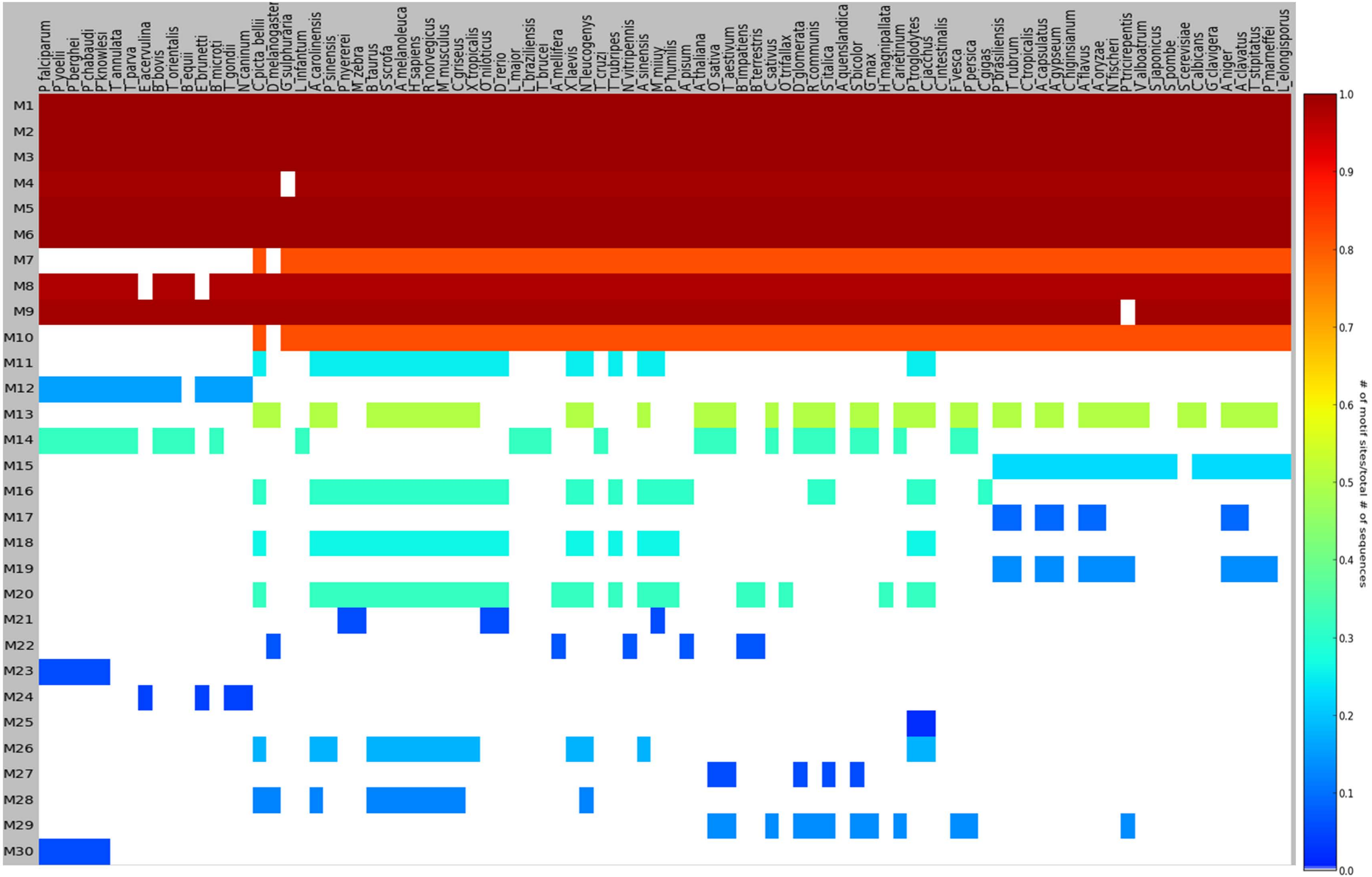

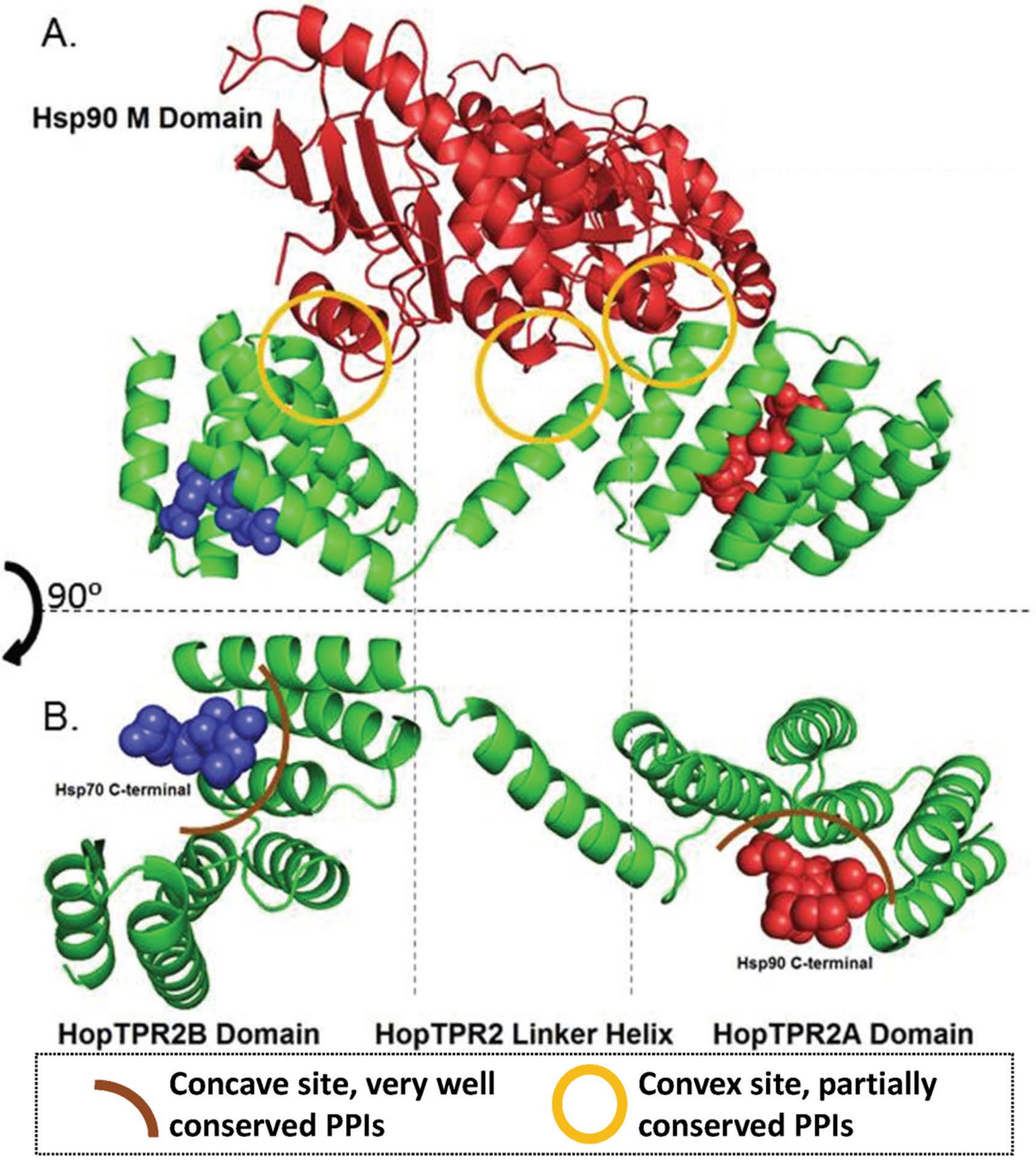

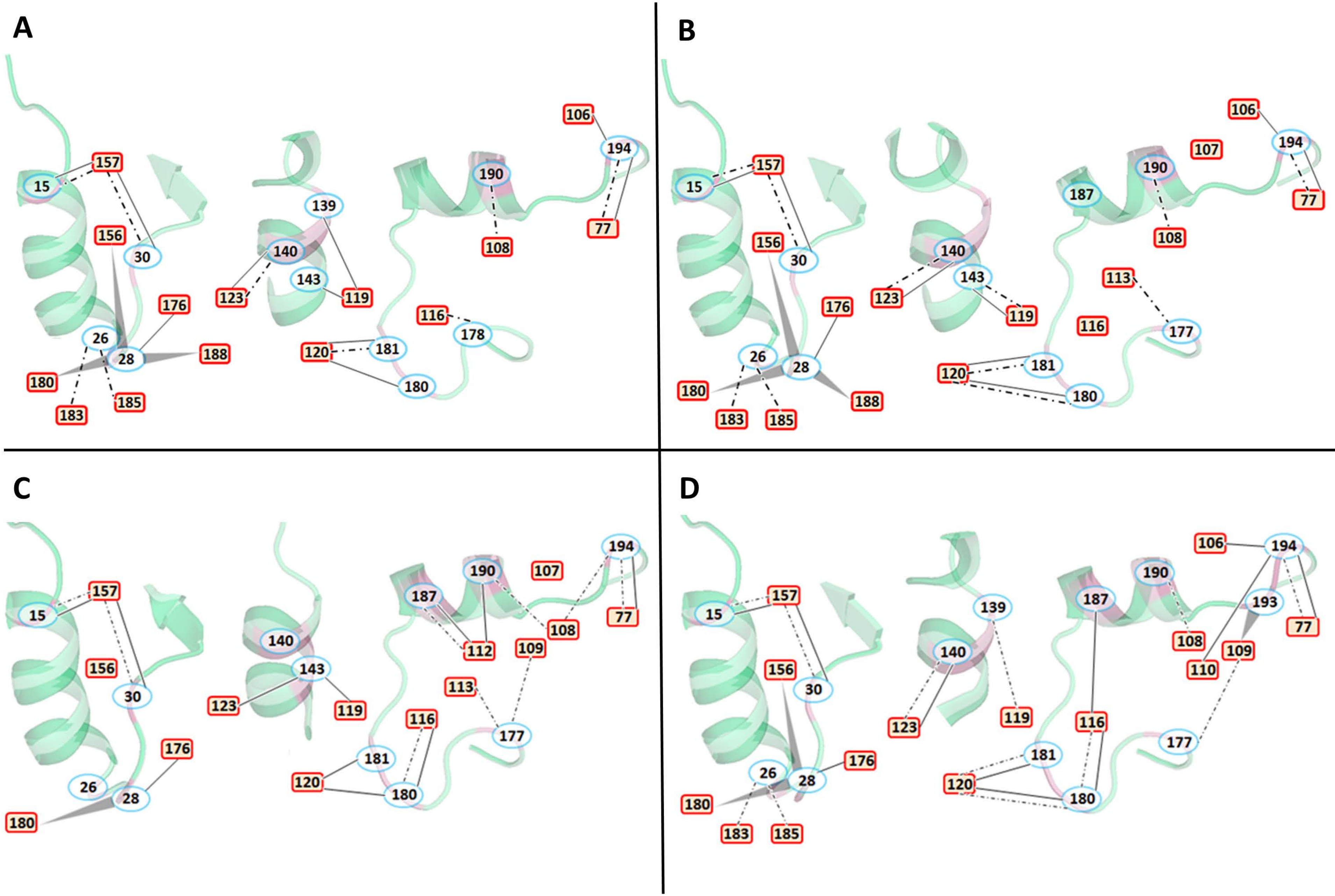

## Supplementary Data 1:

**List of PfHop homolog proteins used in the alignments, phylogenetic tree calculations and motif (MEME) analysis.** It includes the organism that the protein comes from, description of the protein, protein and gene ID reference numbers and some of the BLAST scores (E-value, bit scores, sequence identity, positives and gaps).

## Supplementary Data 2:

**MEME analysis up to 30 motifs in Hop protein sequences.** The logos for motifs displayed as block diagrams. Meme Parameters: Full-length protein, motif width 2-150, multiple occurrences per sequence, all other parameters = default. Significant Motif Overlap: MAST analysis indicated that motif 5 share a similarity with motif 1 and 8 of 0.62. **A)** Location of the motifs, **B)** Motifs and regular expressions. **C)** Mapping of TPR2 motifs to 3D structures at the Hsp90MC-HopTPR2AB interface. i) The positions of the Hop motifs are shown as they occur on the TPR2AB domains of both HsHop and PfHop. ii) The motifs shown in (i) are mapped to a model of the PfHsp90MC-PfHopTPR2AB original orientation. PfHsp90 is colored red, PfHop is colored dark blue and the motifs are colored as in (i). Residues which are identical in both PfHop and HsHop are colored orange in (iii). iv) The model shown in (iii) is rotated 180° and a white circle indicates a region of the interaction interface where there are no conserved residues between human and P. falciparum.

## Supplementary Data 3:

**Summary of templates and models used in this study.** The name of each structure is given along with the calculated DOPE Z-score and MetaMQAPII predicted GDT_TS score. For complexes docked using HADDOCK, the HADDOCK score for the cluster the complex was grouped into by the HADOOCK webserver is given.

A) Structures used for the analysis of the interaction between the convex region of the Hop TPR2A/TPR2B domains and the Hsp90 M domain. For the complexes, the following prefixes are used: Alpha – HsHsp90-alpha-HsHop; Beta – HsHsp90-beta-HsHop; Pf – PfHsp90-PfHop. The rest of the complex name relates to the conditions used to create the complex.

B) Structures used for the analysis of the interaction between the concave regions of the Hop TPR1, TPR2A and TPR2B domains with a short C-terminal peptide from Hsp70 or Hsp90. A description is given above each set of models, including which Hop TPR domain is interacting, as well as the sequence of the short C-terminal Hsp70/Hsp90 peptide, shown in subscript.

## Supplementary Data 4:

**A) Residue mapping from sequences used to models created.** The numbering of residues reported in this study is shown in the 'modeled' column. The actual residues and residue numbers, as they appear in their respective full-length sequences is shown in the adjacent columns.

**B) Summary of interactions displayed in Fig. 4.** Interactions are listed as Hsp90 residues interacting with Hop residues respectively, with interaction types shown in square brackets as follows: I – ionic interaction; H – hydrogen bond; P – hydrophobic interaction. The conserved column represents interactions seen in all three complex sets. The other columns are named as follows: alpha – HsHsp90-alpha-HsHop; beta – HsHsp90-beta-HsHop; template – Schmid *et al.* template (ScHsp90-ScHop). Residue numbers are shown based on their positions in the models created, to make them comparable. For actual residue numbers, refer to (A).

**C) Computational Alanine scanning results.** Each column represents a different group of complexes, named as in (B). Residues listed are hot spot residues, with a $\Delta\Delta G \geq 1 kcal.mol^{-1}$ for more than half of the complexes in the group. Residue numbers are shown based on their positions in the models created, to make them comparable. Actual residue numbers are displayed in brackets.

## Supplementary Data 5:

**A) Summary of interactions between HopTPR concave regions and Hsp70/Hsp90 C-terminal regions.** Interactions are listed as Hop TPR residues interacting with Hsp70 or Hsp90 C-terminal residues, respectively. Interaction types are displayed as in S-Data 4B. Hop TPR residues are numbered according to their positions in the TPR domains (refer to C). Each column represents interactions that were found in both the human and *P. falciparum* models (left), in the human models only (center) or in the *P. falciparum* models only (right).

**B) Comparison of conserved interactions found in each TPR domain.** Each of the interfaces represented (A) is shown, with only interactions common to both human and *P. falciparum* displayed, with interactions aligned for comparison. *In TPR2A, residues K77 and R81 correspond to K70 and R74, respectively in the other TPR domains. This is due to a seven-residue gap present when aligning these sequences as seen in (C)**.**

**C) Alignment of TPR domains used in this study.** TPR1, TPR2A and TPR2B domains are aligned to better illustrate the conservation of interactions among these different domains. Residues which interacted with the Hsp70/Hsp90 C-terminus in each complex are highlighted in green. TPR2A and TPR2B include *S. cerevisiae* sequences as these were present in both template 3UQ3 and 3UPV. Template 1ELW consisted of human sequences, so no additional sequence was required for TPR1.

## Supplementary Data 1:

| Species *(length)* | Protein ID | E-value | Score | % Identity | Positives | Gaps | Gene ID | Description |
|---|---|---|---|---|---|---|---|---|
| *Acyrthosiphon pisum (565)* | XP_001950745.1 | 4.00E-112 | 348 bits (892) | 204/565 (36%) | 319/565 (56%) | 31/565 (5%) | GENE ID: 100167947 LOC100167947 | PREDICTED: stress-induced-phosphoprotein 1-like |
| *Ailuropoda melanoleuca (565)* | XP_002916718.1 | 8.00E-108 | 337 bits (864) | 214/565 (38%) | 323/565 (57%) | 28/565 (5%) | GENE ID: 100475133 LOC100475133 | PREDICTED: stress-induced-phosphoprotein 1-like |
| *Ajellomyces capsulatus NAm1 (574)* | XP_001540631.1 | 1.00E-91 | 296 bits(757 | 196/575(34%) | 300/575(52%) | 22/575(3%) | Gene ID: 544717 | hypothetical protein |
| *Alligator sinensis (560)* | XP_006036942.1 | 6.00E-110 | 343 bits(880 | 212/564(38%) | 318/564(56%) | 36/564(6%) | Gene ID: 102373864 | stress-induced-phosphoprotein 1 |
| *Amphimedon queenslandica (554)* | XP_003387638.1 | 8.00E-101 | 318 bits (815) | 198/564 (35%) | 319/564 (57%) | 20/564 (4%) | GENE ID: 100633434 LOC100633434 | PREDICTED: Stress-induced phosphoprotein 1-like |
| *Anolis carolinensis (566)* | XP_003228007.1 | 1.00E-112 | 349 bits (896) | 216/566 (38%) | 327/566 (58%) | 30/566 (5%) | GENE ID: 100563364 LOC100563364 | PREDICTED: stress-induced-phosphoprotein 1-like |
| *Apis mellifera (539)* | XP_006567267.1 | 4.00E-115 | 356 bits(913) | 213/562(38%) | 322/562(57%) | 38/562(6%) | Gene ID: 102656934 | stress-induced-phosphoprotein 1-like |
| *Arabidopsis thaliana (558)* | NP_001031620.1 | 5.00E-121 | 372 bits (954) | 218/572 (38%) | 334/572 (58%) | 29/572 (5%) | GENE ID: 826849 AT4G12400 | putative stress-inducible protein |
| *Arthroderma gypseum [CBS 118893] (578)* | XP_003175251.1 | 5.00E-89 | 288 bits (736) | 202/589 (34%) | 303/589 (51%) | 45/589 (8%) | | heat shock protein STI1 |
| *Aspergillus clavatus [NRRL 1] (581)* | XP_001272361.1 | 1.00E-81 | 268 bits (686) | 188/585 (32%) | 297/585 (51%) | 35/585 (6%) | GENE ID: 4704565 ACLA_065690 | heat shock protein (Sti1), putative |
| *Aspergillus flavus [NRRL3357] (579)* | XP_002380660.1 | 1.00E-87 | 285 bits (728) | 195/584 (33%) | 300/584 (51%) | 35/584 (6%) | GENE ID: 7914463 AFLA_071010 | heat shock protein (Sti1), putative |
| *Aspergillus niger [CBS 513.88] (580)* | XP_001395168.2 | 1.00E-83 | 273 bits (699) | 187/584 (32%) | 304/584 (52%) | 34/584 (6%) | GENE ID: 4985429 ANI_1_116104 | heat shock protein STI1 |
| *Aspergillus oryzae [RIB40] (579)* | XP_001825463.1 | 9.00E-87 | 285 bits (728) | 195/584 (33%) | 300/584 (51%) | 35/584 (6%) | GENE ID: 5997558 AOR_1_950074 | shock protein STI1 |
| *Babesia bovis (546)* | XP_001611358.1 | 0 | 546 bits (1407) | 273/553 (49%) | 394/553 (71%) | 12/553 (2%) | GENE ID: 5479603 BBOV_III002230 | tetratricopeptide repeat domain containing protein |
| *Babesia equi (560)* | XP_004832859.1 | 0 | 527 bits(1358) | 276/553(50%) | 382/553(69%) | 18/553(3%) | Gene ID: 15803014 | hypothetical protein |
| *Babesia microti (547)* | CCF75306.1 | 1.00E-178 | 519 bits(1336) | 267/561(48%) | 377/561(67%) | 24/561(4%) | | |
| *Bombus impatiens (539)* | XP_003486756.1 | 1.00E-114 | 353 bits(907) | 207/555(37%) | 324/555(58%) | 24/555(4%) | Gene ID: 100743215 | stress-induced-phosphoprotein 1-like |
| *Bombus terrestris (565)* | XP_003402501.1 | 8.00E-114 | 352 bits (902) | 207/555 (37%) | 324/555 (58%) | 24/555 (4%) | GENE ID: 100631059 cactus-2 | PREDICTED: hypothetical protein LOC100631059 |
| *Bos taurus (565)* | NP_001030569.1 | 5.00E-107 | 335 bits (859) | 214/565 (38%) | 324/565 (57%) | 28/565 (5%) | GENE ID: 617109 STIP1 | stress-induced-phosphoprotein 1 |
| *Callithrix jacchus (586)* | XP_002755509.1 | 2.00E-107 | 336 bits (861) | 212/565 (38%) | 326/565 (58%) | 28/565 (5%) | GENE ID: 100409178 STIP1 | PREDICTED: stress-induced-phosphoprotein 1 isoform 1 |
| *Candida albicans [SC5314] (590)* | XP_714740.1 | 4.00E-88 | 286 bits (732) | 198/598 (33%) | 304/598 (51% | 51/598 (9%) | GENE ID: 3643631 STI1 | hypothetical protein CaO19.10702 |
| *Candida tropicalis[MYA-3404] (579)* | XP_002551007.1 | 2.00E-97 | 309 bits (792) | 198/585 (34%) | 310/585 (53% | 36/585 (6%) | GENE ID: 8299626 CTRG_05305 | heat shock protein STI1 |
| *Chrysemys picta bellii (545)* | XP_005305601.1 | 7.00E-117 | 361 bits(927) | 221/565(39%) | 330/565(58%) | 26/565(4%) | Gene ID: 101945489 | stress-induced-phosphoprotein 1 |
| *Cicer arietinum (582)* | XP_004511659.1 | 6.00E-113 | 352 bits(902) | 212/590(36%) | 319/590(54% | 41/590(6%) | Gene ID: 101494428 LOC101494428 | heat shock protein STI-like |
| *Ciona intestinalis (570)* | XP_002128875.1 | 2.00E-100 | 317 bits (813) | 197/570 (35%) | 314/570 (55%) | 37/570 (6%) | GENE ID: 100181490 LOC100181490 | PREDICTED: similar to Stress-induced-phosphoprotein 1 (STI1) |
| *Colletotrichum higginsianum (580)* | CCF38418.1 | 5.00E-64 | 221 bits(563) | 128/320(40%) | 193/320(60%) | 8/320(2%) | | tetratricopeptide |
| *Crassostrea gigas (546)* | EKC18743.1 | 7.00E-105 | 330 bits(846) | 198/577(34%) | 327/577(56%) | 30/577(5%) | | Stress-induced-phosphoprotein 1 |
| *Cricetulus griseus (565)* | NP_001233607.1 | 2.00E-108 | 338 bits (868) | 214/565 (38% | 325/565 (58%) | 28/565 (5%) | GENE ID: 100689413 Stip1 | stress-induced-phosphoprotein 1 |
| *Cucumis sativus (577)* | XP_004147938.1 | 6.00E-119 | 367 bits(942) | 215/585(37%) | 332/585(56%) | 36/585(6%) | Gene ID: 101221871 | heat shock protein STI-like |
| *Danio rerio (565)* | NP_001007767.1 | 3.00E-109 | 341 bits (874) | 211/565 (37%) | 325/565 (58%) | 29/565 (5%) | GENE ID: 493606 stip1 | stress-induced-phosphoprotein 1 |
| *Drosophila melanogaster (490)* | NP_477354.1 | | 1.00E-74 | 188/490 (38%) | 278/490 (56%) | 24/490(4%) | GENE ID: 33202 Hop | Hsp70/Hsp90 organizing protein homolog |
| *Eimeria acervulina (551)* | CDI79732.1 | 0 | 548 bits(1412) | 283/563(50%) | 378/563(67%) | 16/563(2%) | | Hsc70/Hsp90-organizing protein, putative |
| *Eimeria brunetti (554)* | CDJ49419.1 | 0 | 530 bits(1366) | 270/563(48%) | 382/563(67%) | 13/563(2%) | | Hsc70/Hsp90-organizing protein, putative |
| *Fragaria vesca subsp. Vesca (586)* | XP_004298670.1 | 4.00E-115 | 357 bits(916) | 211/596(35%) | 333/596(55%) | 49/596(8%) | Gene ID: 101293091 | heat shock protein STI-like |
| *Galdieria sulphuraria (571)* | XP_005705574.1 | 4.00E-119 | 366 bits(939) | 224/581(39%) | 346/581(59%) | 29/581(4%) | Gene ID: 17087882 | stress-induced-phosphoprotein 1 |
| *Giardia lamblia ATCC 50803* | XP_001704357.1 | 6.00E-38 | 147 bits(371) | 111/357(31%) | 189/357(52%) | 35/357(9%) | Gene ID: 5697231 GL50803_27310 | Stress-induced-phosphoprotein 1 |
| *Glycine max (585)* | XP_003538668.1 | 2.00E-114 | 357 bits(915) | 215/591(36%) | 326/591(55%) | 40/591(6%) | Gene ID: 100816776 | heat shock protein STI-like |
| *Grosmannia clavigera kw1407 (603)* | EFX03578.1 | 2.00E-24 | 105 bits(263) | 53/112(47%) | 72/112(64%) | 0/112(0%) | | heat shock protein sti1 |
| *Homo sapiens (565)* | NP_006810.1 | 2.00E-106 | 333 bits (855) | 212/565 (38%) | 326/565 (58%) | 28/565 (5%) | GENE ID: 10963 STIP1 | stress-induced-phosphoprotein 1 |
| *Hydra magnipapillata (534)* | XP_002160503.2 | 8.00E-97 | 307 bits (787) | 195/560 (35%) | 308/560 (55%) | 28/560 (5%) | GENE ID: 100203295 LOC100203295 | PREDICTED: similar to stress-induced-phosphoprotein 1 (Hsp70/Hsp90-organizing |
| *Leishmania braziliensis [MHOM/BR/75/M2904] (547)* | XP_001562145.1 | 9.00E-110 | 341 bits (875) | 209/563 (37%) | 320/563 (57%) | 26/563 (5%) | GENE ID: 5413050 LBRM_08_0880 | stress-induced protein sti1 |
| *Leishmania infantum [JPCM5] (546)* | XP_001463435.1 | 2.00E-118 | 363 bits (933) | 215/563 (38%) | 326/563 (58%) | 27/563 (5%) | GENE ID: 5066714 LINJ_08_1020 | stress-induced protein sti1 |
| *Leishmania major [strain Friedlin] (546)* | XP_001681140.1 | 7.00E-118 | 362 bits (929) | 214/563 (38%) | 326/563 (58%) | 27/563 (5%) | GENE ID: 5649395 LMJF_08_1110 | stress-induced protein sti1 |
| *Lodderomyces elongisporus [NRRL YB-4239] (596)* | XP_001524727.1 | 8.00E-88 | 285 bits (728) | 196/600 (33%) | 314/600 (52%) | 49/600 (8%) | GENE ID: 5232040 LELG_03759 | heat shock protein STI1 |
| *Maylandia zebra (542)* | XP_004557346.1 | 2.00E-110 | 344 bits(883) | 328/571(57%) | 328/571(57%) | 41/571(7%) | Gene ID: 101463850 LOC101463850 | stress-induced-phosphoprotein 1-like |
| *Miichthys miiuy (542)* | ADP05116.1 | 2.00E-112 | 342 bits(878) | 212/569(37%) | 331/569(58%) | 37/569(6%) | | stress-induced-phosphoprotein 1 |
| *Mus musculus (565)* | NP_058017.1 | 3.00E-108 | 338 bits (868 | 214/565 (38%) | 326/565 (58%) | 28/565 (5%) | GENE ID: 20867 Stip1 | stress-induced-phosphoprotein 1 |
| *Nasonia vitripennis (565)* | XP_001603429.1 | 1.00E-106 | 333 bits (855) | 204/569 (36%) | 323/569 (57%) | 32/569 (6%) | GENE ID: 100119701 LOC100119701 | PREDICTED: stress-induced-phosphoprotein 1-like |
| *Neosartorya fischeri [NRRL 181] (582)* | XP_001262823.1 | 1.00E-90 | 292 bits (747) | 191/582 (33%) | 299/582 (51%) | 29/582 (5%) | GENE ID: 4589462 NFIA_114590 | heat shock protein (Sti1), putative |
| *Neospora caninum [Liverpool] (563)* | XP_003880293.1 | 2.00E-168 | 493 bits (1269) | 268/566 (47%) | 370/566 (65%) | 7/566 (1%) | GENE ID: 13446323 NCLIV_007330 | similar to uniprot\|P15705 Saccharomyces cerevisiae YOR027w STI1, related |
| *Nomascus leucogenys (543)* | XP_004093067.1 | 8.00E-105 | 329 bits(844) | 210/565(37%) | 324/565(57%) | 28/565(4%) | Gene ID: 100597805 STIP1 | stress-induced-phosphoprotein 1 |
| *Oreochromis niloticus (571)* | XP_003450486.1 | 8.00E-111 | 345 bits (884) | 215/571 (38%) | 328/571 (57%) | 41/571 (7%) | GENE ID: 100696373 LOC100696373 | stress-induced-phosphoprotein 1-like |
| *Oryza sativa [Japonica Group] (578)* | NP_001047563.1 | 8.00E-116 | 358 bits (920) | 213/600 (36%) | 328/600 (55%) | 65/600 (11%) | GENE ID: 4330134 Os02g0644100 | Os02g0644100 |
| *Oxytricha trifallax (566)* | EJY72317.1 | 3.00E-117 | 362 bits(930) | 209/575(36%) | 335/575(58%) | 36/575(6%) | | TPR repeat-containing protein |
| *Pan troglodytes (560)* | NP_001267378.1 | 3.00E-105 | 332 bits (851) | 211/560 (38%) | 324/560 (58%) | 28/560 (5%) | GENE ID: 451286 STIP1 | PREDICTED: stress-induced-phosphoprotein 1 isoform 1 |
| *Pelodiscus sinensis (545)* | XP_006129808.1 | 2.00E-113 | 352 bits(903 | 215/560(38%) | 328/560(58%) | 28/560(5%) | Gene ID: 102450034 | stress-induced-phosphoprotein 1 |
| *Penicilium marneffei (578)* | XP_002146473.1 | 4.00E-84 | 275 bits (703) | 182/590 (31%) | 297/590 (50%) | 48/590 (8%) | GENE ID: 7024148 PMAA_070140 | heat shock protein (Sti1), putative |
| *Plasmodium berghei [strain ANKA] (559)* | XP_677465.1 | 0 | 880 bits (2275) | 462/564 (82%) | 519/564 (92%) | 5/564 (1%) | GENE ID: 3426000 PB000909.03.0 | hypothetical protein |
| *Plasmodium chabaudi chabaudi (559)* | XP_745506.1 | 0 | 880 bits (2273) | 460/564 (82%) | 519/564 (92%) | 5/564 (1%) | GENE ID: 3498629 PC000814.02.0 | hypothetical protein |
| *Plasmodium falciparum [3D7] (564)* | PF3D7_1434300 and XP_001348498.1 | 0 | 1144 bits (2959) | 564/564 (100%) | 564/564 (100%) | 0/564 (0%) | GENE ID: 811906 PF14_0324 | Hsp70/Hsp90 organizing protein, putative |
| *Plasmodium knowlesi [strain H] (560)* | XP_002260669.1 | 0 | 885 bits (2287) | 460/564 (82%) | 509/564 (90%) | 4/564 (1%) | GENE ID: 7322649 PKH_131500 | hypothetical protein, conserved in Apicomplexan species |
| *Plasmodium yoelii yoelii [17XNL] (559)* | XP_731105.1 | 0 | 882 bits (2278 | 463/564 (82%) | 520/564 (92%) | 5/564 (1%) | GENE ID: 3830331 PY03138 | stress-induced protein Sti1 |
| *Prunus persica (573)* | EMJ11701.1 | 1.00E-116 | 360 bits(925) | 210/582(36%) | 324/582(55%) | 34/582(5%) | | hypothetical protein PRUPE_ppa003460mg |
| *Pseudopodoces humilis (541)* | XP_005533270.1 | 3.00E-109 | 340 bits(873) | 202/565(36%) | 312/565(55%) | 30/565(5%) | Gene ID: 102101929 STIP1 | stress-induced-phosphoprotein 1 |
| *Pundamilia nyererei (541)* | XP_005749595.1 | 5.00E-112 | 348 bits(894) | 212/565(38%) | 329/565(58%) | 30/565(5%) | Gene ID: 102214293 | stress-induced-phosphoprotein 1-like |
| *Pyrenophora tritici-repentis [Pt-1C-BFP] (576)* | XP_001940455.1 | 6.00E-87 | 282 bits (722) | 189/581 (33%) | 308/581 (53%) | 32/581 (6%) | GENE ID: 6348424 PTRG_10123 | heat shock protein STI1 |
| *Rattus norvegicus (565)* | NP_620266.1 | 1.00E-107 | 337 bits (864) | 212/565 (38%) | 327/565 (58% | 28/565 (5%) | GENE ID: 192277 Stip1 | stress-induced-phosphoprotein 1 |
| *Ricinus communis (578)* | XP_002509580.1 | 2.00E-117 | 363 bits(931) | 211/591(36%) | 322/591(54%) | 47/591(7%) | Gene ID: 8272118 RCOM_1679700 | heat shock protein 70 (HSP70)-interacting protein, putative |
| *Saccharomyces cerevisiae [S288c] (589)* | NP_014670.1 | 2.00E-97 | 310 bits (793) | 203/593 (34%) | 316/593 (53%) | 46/593 (8%) | GENE ID: 854192 STI1 | Sti1p |
| *Saccoglossus kowalevskii (310)* | XP_002733893.1 | 3.00E-75 | 244 bits (622) | 127/315 (40%) | 203/315 (64%) | 8/315 (3%) | GENE ID: 100374768 LOC100374768 | PREDICTED: stress-induced-phosphoprotein 1 (Hsp70/Hsp90-organizing |
| *Schizosaccharomyces japonicas [yFS275] (582)* | XP_002174852.1 | 1.00E-96 | 305 bits (781) | 200/585 (34%) | 313/585 (54%) | 33/585 (6%) | GENE ID: 7052233 SJAG_03717 | chaperone activator Sti1 |
| *Schizosaccharomyces pombe [972h-] (591)* | NP_588123.1 | 3.00E-92 | 296 bits (757) | 202/596 (34%) | 302/596 (51%) | 46/596 (8%) | GENE ID: 2539474 sti1 | chaperone activator Sti1 (predicted) |
| *Setaria italica (580)* | XP_004953274.1 | 4.00E-117 | 362 bits(930) | 214/595(36%) | 330/595(55%) | 53/595(8%) | Gene ID: 101758693 LOC101758693 | heat shock protein STI-like |
| *Sorghum bicolor (580)* | XP_002446861.1 | 1.00E-118 | 366 bits(940) | 210/587(36%) | 328/587(55%) | 37/587(6%) | Gene ID: 8058346 | hypothetical protein |
| *Sus scrofa (565)* | XP_003353842.1 | 3.00E-108 | 338 bits (866) | 216/565 (38%) | 323/565 (57%) | 28/565 (5%) | GENE ID: 100623923 LOC100623923 | PREDICTED: stress-induced-phosphoprotein 1-like |
| *Takifugu rubripes (539)* | XP_003976325.1 | 9.00E-110 | 342 bits(877) | 206/567(36%) | 327/567(57%) | 36/567(6%) | Gene ID: 101066391 | stress-induced-phosphoprotein 1-like |
| *Talaromyces stipitatus [ATCC 10500] (577)* | XP_002478770.1 | 7.00E-83 | 272 bits (695) | 184/586 (31%) | 296/586 (51%) | 41/586 (7%) | GENE ID: 8108134 TSTA_090460 | heat shock protein (Sti1), putative |
| *Theileria annulata [strain Ankara] (540)* | XP_955292.1 | 0 | 543 bits (1398) | 280/554 (51%) | 393/554 (71%) | 18/554 (3%) | GENE ID: 3865063 TA18515 | hypothetical protein, conserved |
| *Theileria orientalis (557)* | BAM40980.1 | 0 | 540 bits(1392) | 276/550(50%) | 383/550(69%) | 18/550(3%) | | |
| *Theileria parva [strain Muguga] (540)* | XP_763615.1 | 0 | 532 bits (1371) | 286/554 (52%) | 393/554 (71%) | 18/554 (3%) | GENE ID: 3499840 TP03_0587 | hypothetical protein |
| *Toxoplasma gondii (565)* | ESS30241.1 | 3.00E-168 | 494 bits(1272) | 269/566(48%) | 366/566(64%) | 5/566(0%) | | tetratricopeptide repeat domain containing protein |
| *Trichophyton rubrum [CBS 118892] (578)* | XP_003232880.1 | 2.00E-88 | 286 bits (731) | 202/585 (35%) | 299/585 (51%) | 37/585 (6%) | GENE ID: 10372247 TERG_06870 | heat shock protein STI1 |
| *Triticum aestivum (581)* | ADN05856.1 | 3.00E-116 | 358 bits(919) | 216/590(37%) | 329/590(55%) | 42/590(7%) | Gene ID: 100682494 | HOP |
| *Trypanosoma bruceibrucei [strain 927/4 GUTat10.1] (550)* | XP_844966.1 | 4.00E-110 | 342 bits (877) | 205/562 (36%) | 316/562 (56%) | 21/562 (4%) | GENE ID: 3657403 Tb927.5.2940 | stress-induced protein sti1 |
| *Trypanosoma cruzi (556)* | ESS64435.1 | 2.00E-111 | 348 bits(892) | 209/568(37%) | 324/568(57%) | 27/568(4%) | | stress-induced protein sti1 |
| *Verticillium albo-atrum [VaMs.102] (584)* | XP_002999908.1 | 4.00E-87 | 283 bits (724) | 195/586 (33%) | 296/586 (51%) | 36/586 (6%) | GENE ID: 9531727 VDBG_09948 | heat shock protein STI1 |
| *Xenopus (Silurana) tropicalis (573)* | NP_989360.1 | 3.00E-108 | 338 bits (867) | 217/573 (38%) | 325/573 (57%) | 44/573 (8%) | GENE ID: 394990 stip1 | stress-induced-phosphoprotein 1 |

## Supplementary Data 2A:

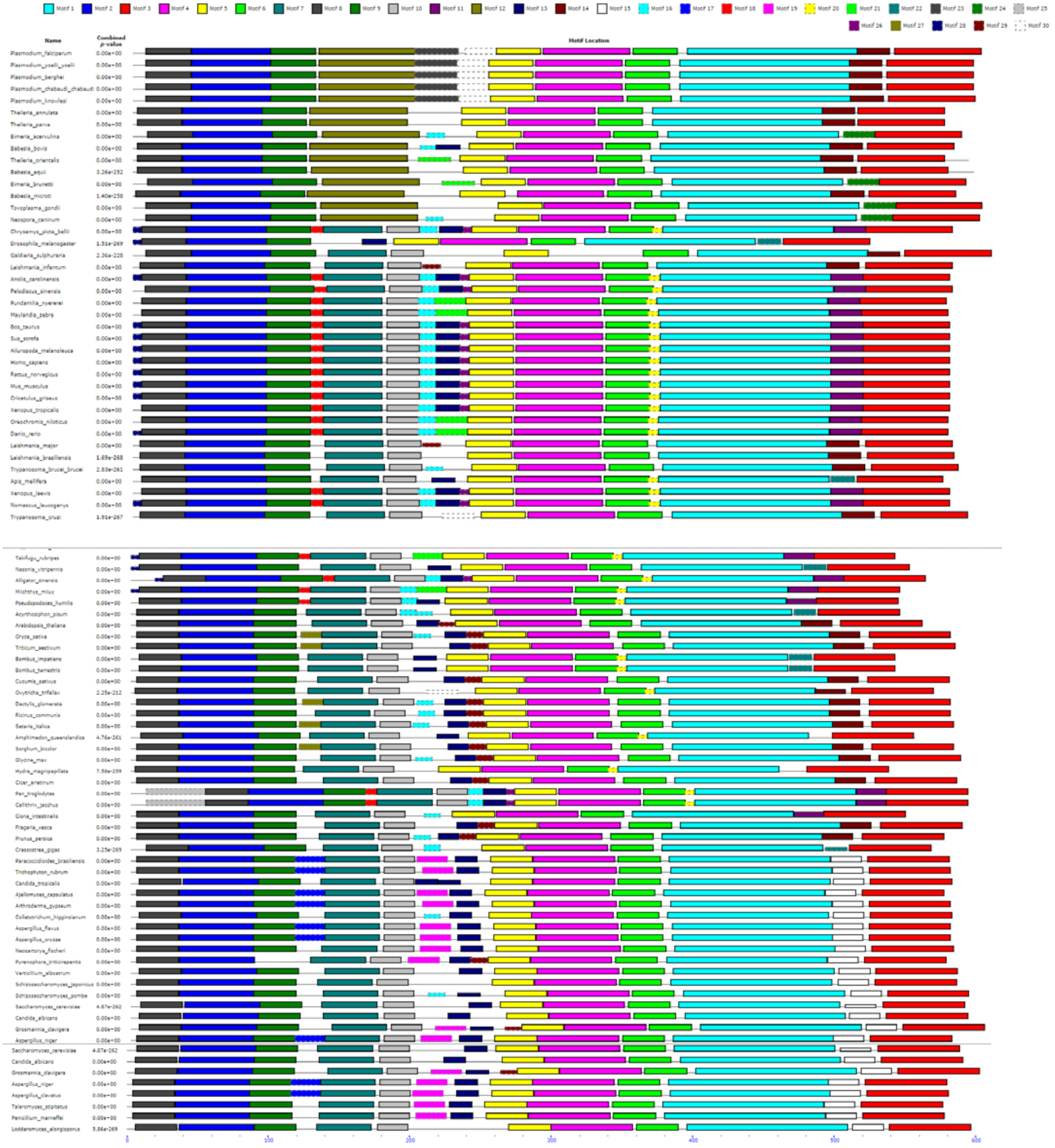

**Supplementary Data 2B:**

## Motif 1

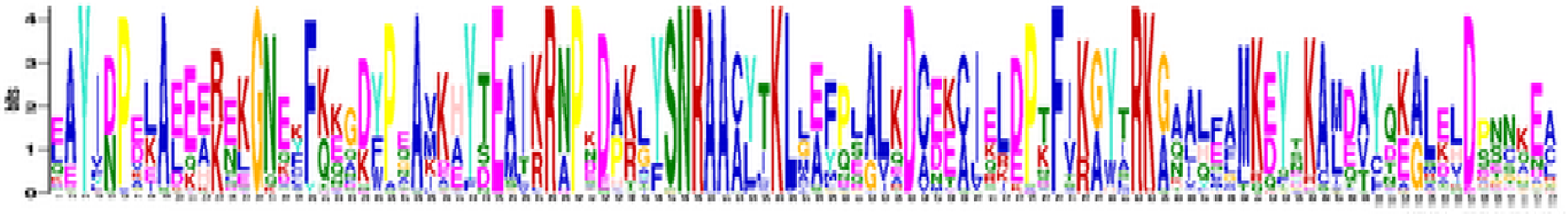

Regex:
[EL]AYI[DN]P[ED][LK]A[EL]E[EA][RK][EN]KGNE[KFY]F[KQ][KE]G[DK][YFW]PEA[VM]K[HAE]Y[TS]E[AM]I[KR]R[NA]P[KN]D[AP][KR][LG]YSNRAA[CA][YL][TI]KL[LG][EA]F[PQ][LS][AG]LKD[CA][ED][KE][CA]IEL[DE]P[TK]F[IV][KR][GA]YTRK[GA][ANQ]AL[FE]AMK[ED]Y[TS]KA[ML][DE][AV]YQ[KE][AG]L[EK][LH]D[PS][NS]NKE[AC]

## Motif 2

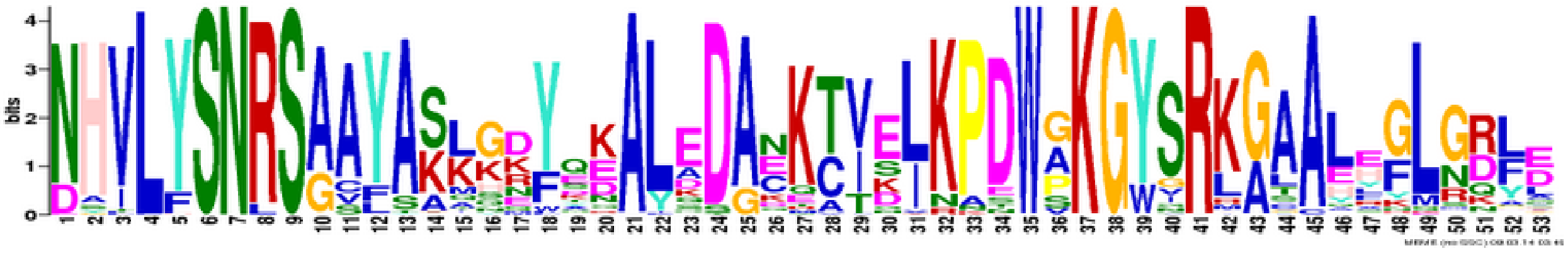

Regex:
NHVLYSNRS[AG]AYA[SK][LK][GK][DK][YF]Q[KE]ALEDA[NCE]K[TC][VI]E[LI]KPDW[GA]KGYSRK[GA]AAL[EH][GF]L[GN][RD][LF][ED]

## Motif 3

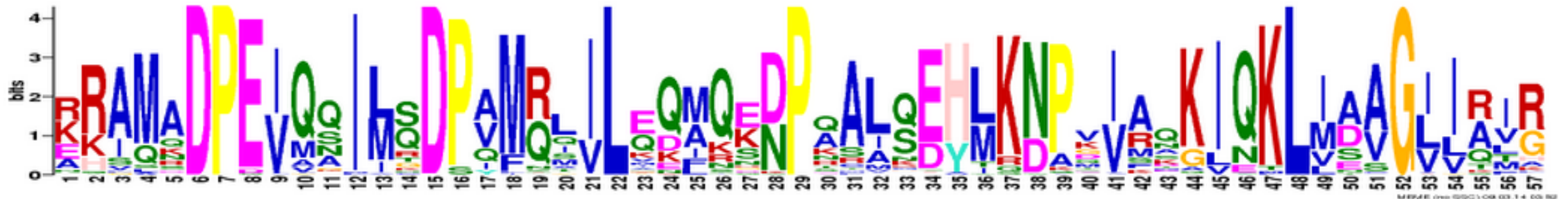

Regex:
[RK]RAMADPE[IV]QQI[LM][SQ]DP[AV]M[RQ]L[IV]L[EQ]Q[MA]Q[EK][DN]P[QA]AL[QS][ED][HY][LM]K[ND]PV[IV]AQKIQKL[IM][AD][AV]G[ILV]I[RA]I[RG]

**Motif 4**

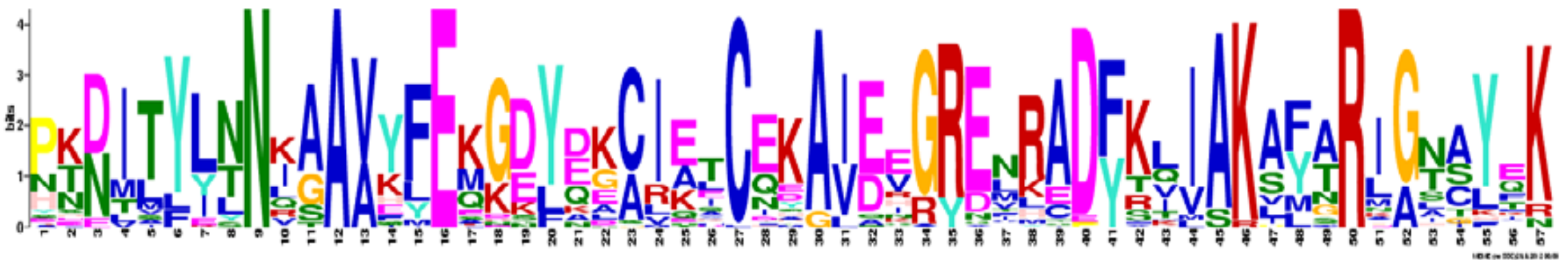

Regex:
P[TK][DN][IM]TY[LI][TN]N[QK]AA[VA]YFE[KM]G[DE]Y[DE]K[CA][IR][EK][LT]CEKA[IV]E[VE]GRENR[AE]D[FY][KR][QL][IV]AKA[YFL][AT]RI[GA][NT][AS]YFK

**Motif 5**

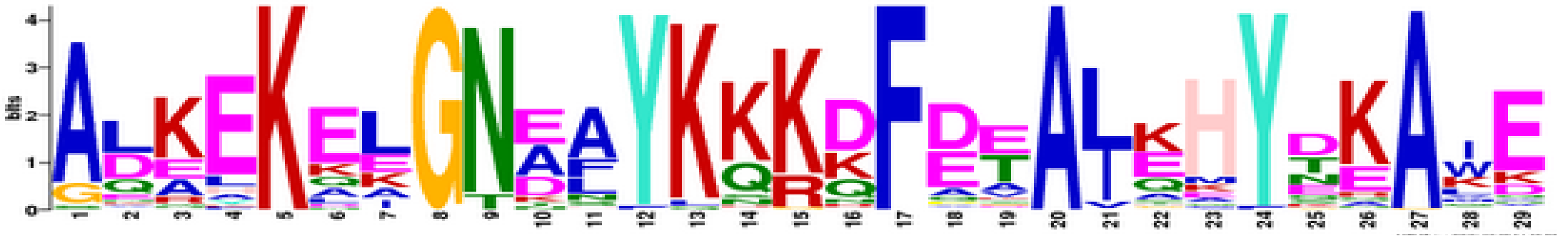

Regex:
A[LD]KEKELGN[EA]AYK[KQ][KR][DK]F[DE][ET]A[LI][KE]HY[DT]KA[IW]E

**Motif 6**

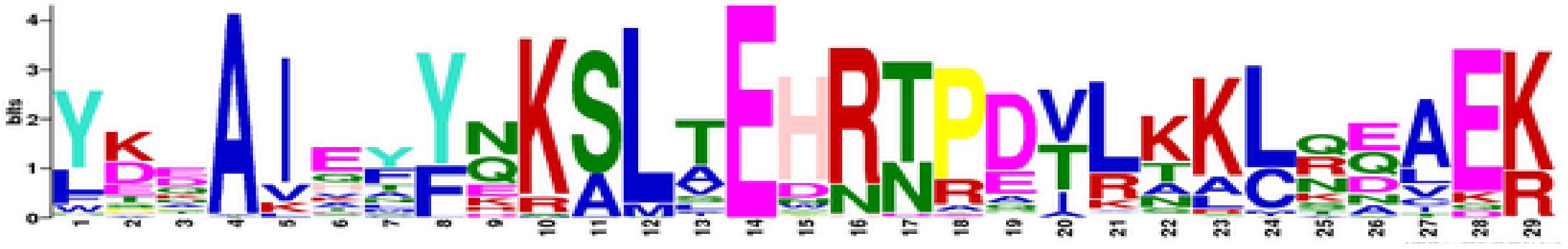

Regex:
Y[KD]xAIE[YF][YF][NQ]K[SA]LTEHR[TN]PD[VT]LKK[LC][QR][EQ]AE[KR]

## Motif 7

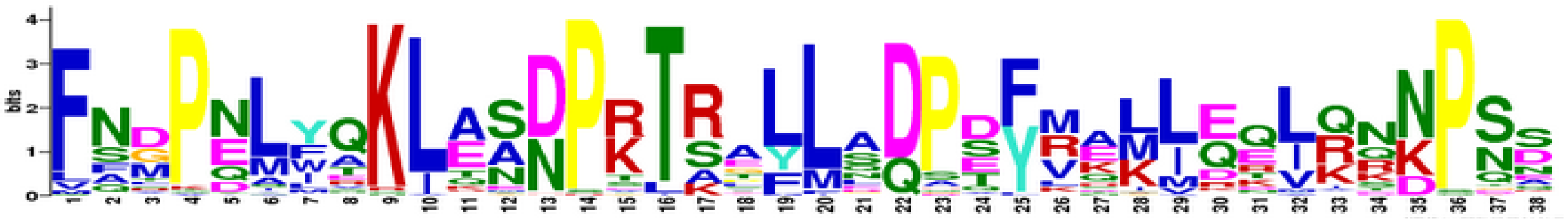

Regex:
FN[DGM]P[NE]LYQKL[AE][SA][DN]P[RK]T[RS]ALL[AS][DQ]P[DS][FY][MRV][AE][LMK][LI][EQ][QE][LI][QRK]N[NK]P[SN][SD]

## Motif 8

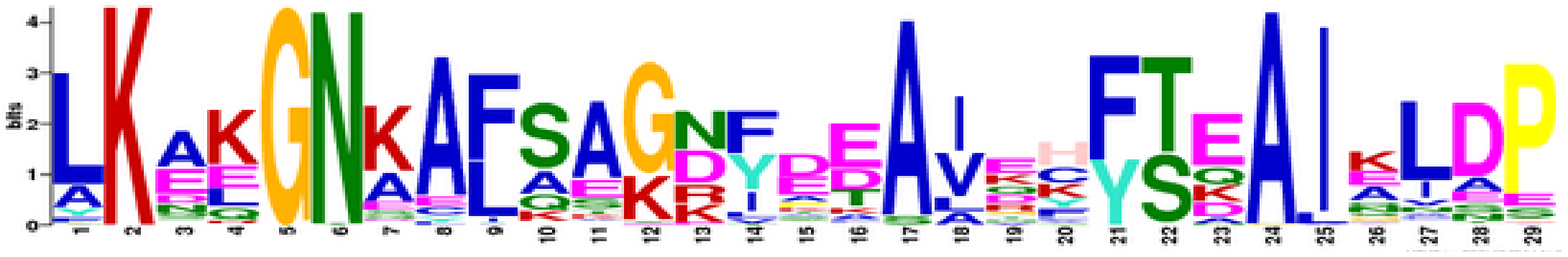

Regex:
LK[AE]KGN[KA]A[FL]SA[GK][ND][FY][DE][ED]A[IV]E[HC][FY][TS]EAI[KE]LDP

## Motif 9

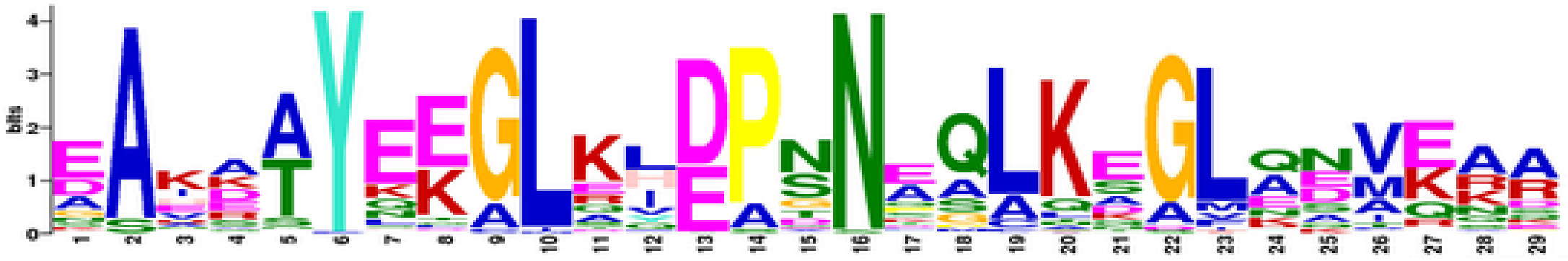

Regex:
EAKA[AT]YE[EK]GLK[LHI][DE]P[NS]N[EA]QLKEGL[QA]NV[EK]A[AR]

## Motif 10

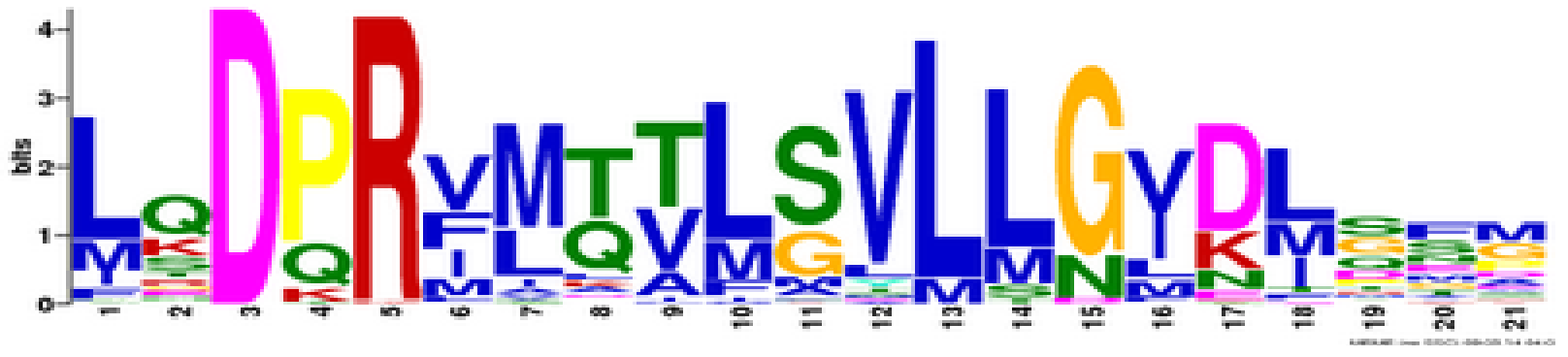

Regex:
LQD[PQ]R[VF][ML][TQ][TV]L[SG]VLLG[VI][DK][LM][SG]F[MG]

**Motif 11**

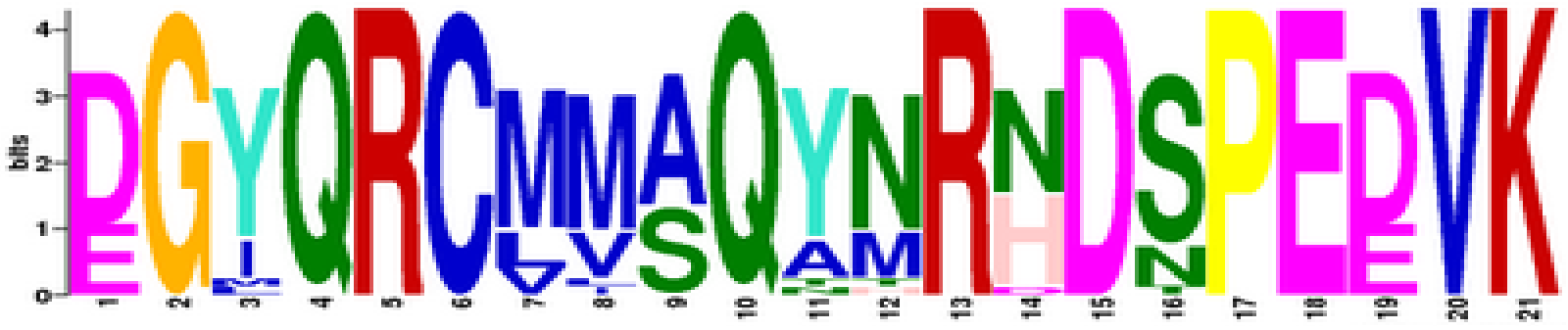

Regex:
[DE]GYQRCM[MV][AS]QY[NM]R[NH]DSPE[DE]VK

**Motif 12**

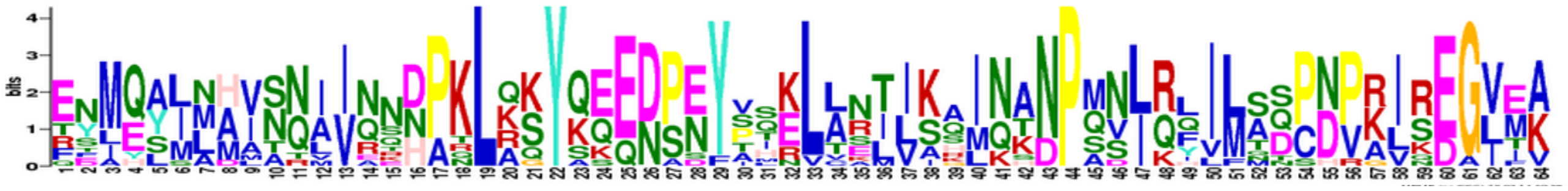

Regex:
E[NY]M[QE][AYS][ILM][LMN][AHM][VI][NS][NQ][IA][IV][NQ]N[DHN][PA]KL[KQAR][KS]Y[QK][EQ][EQ][DN][PS][EN]Y[PSV][QS][KE]L[AL][NR][TIL][ILV][KS][AQ][IM][NQ][AKT][ND]P[MQS][NSV][LI][RQ][LFQ]I[LM][SA][DQS][PC][ND][PV][KRA][IL][RS][ED]G[VLI][EM][AK]

**Motif 13**

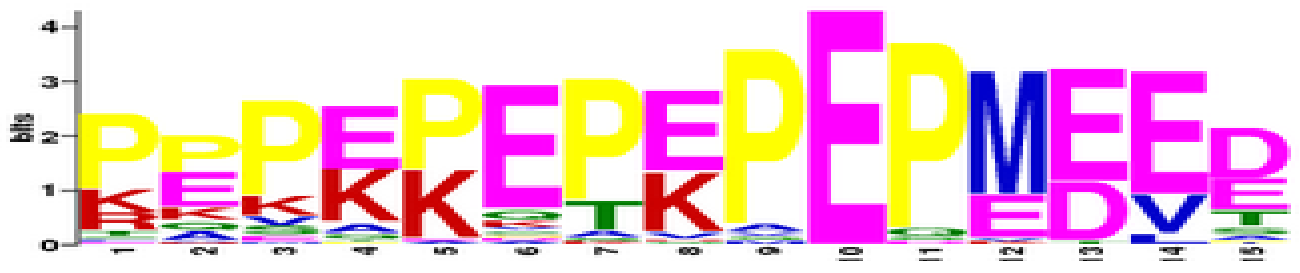

Regex:
P[PE]P[EK][PK]EP[EK]PEP[ME][ED][EV][DE]

**Motif 14**

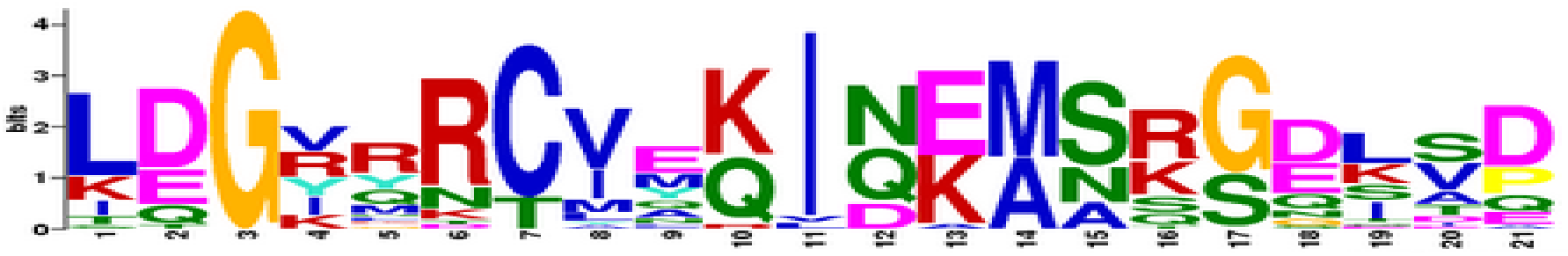

Regex:
L[DE]G[RV]RRC[VI]E[KQ]I[NQ][EK][MA][SN][RK][GS][DE][LK][SV][DP]

---

**Motif 15**

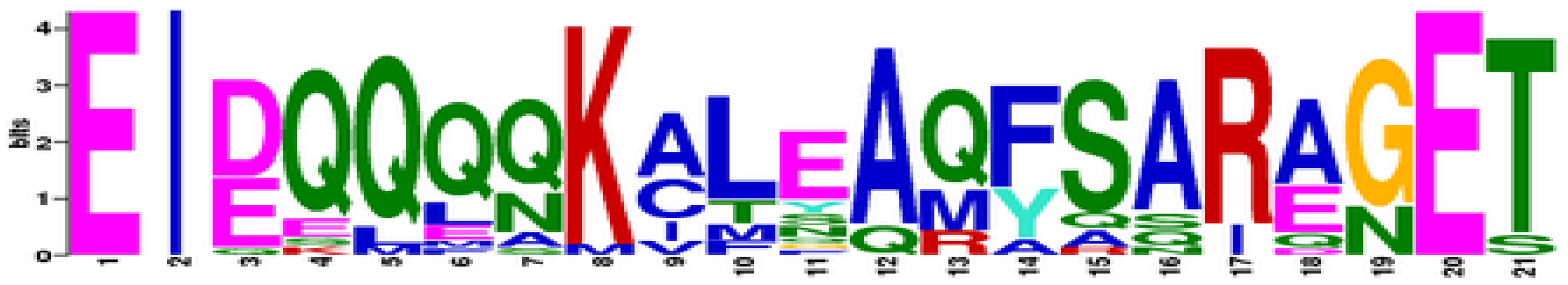

Regex:
EI[DE]QQQ[QN]K[AC]LEA[QM][FY]SAR[AE][GN]ET

---

**Motif 16**

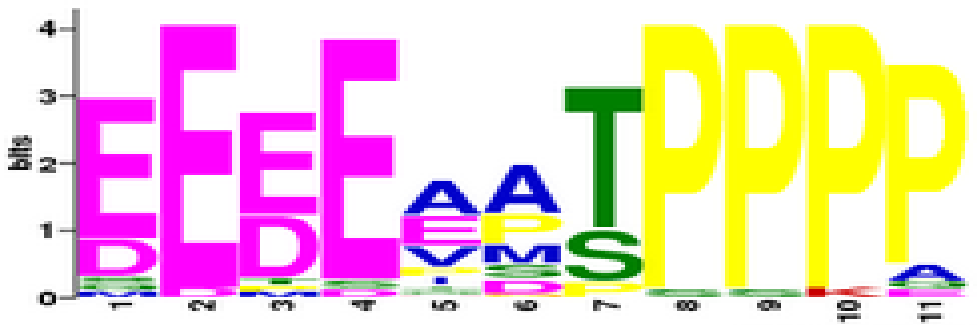

Regex:
EE[ED]E[AE][AP][TS]PPPP

---

**Motif 17**

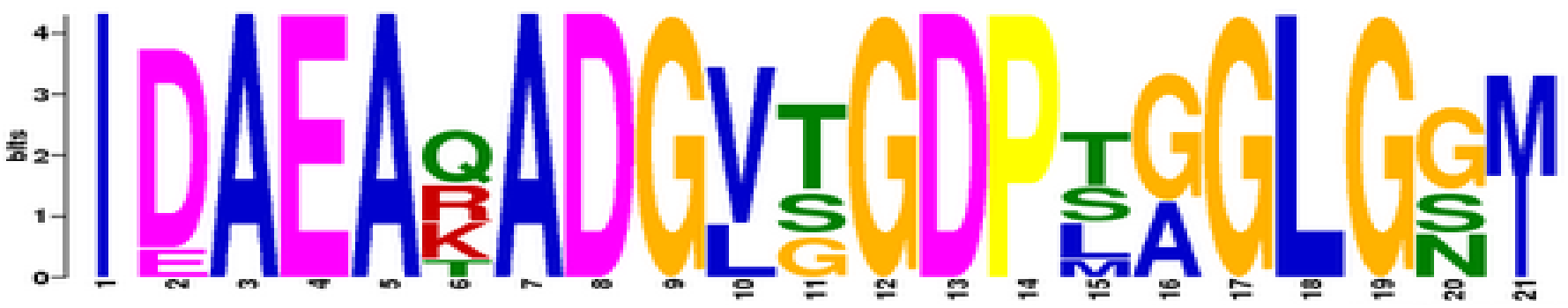

Regex:
IDAEA[QKR]ADG[VL][TGS]GDP[TLS][GA]GLG[GNS][IM]

---

## Motif 18

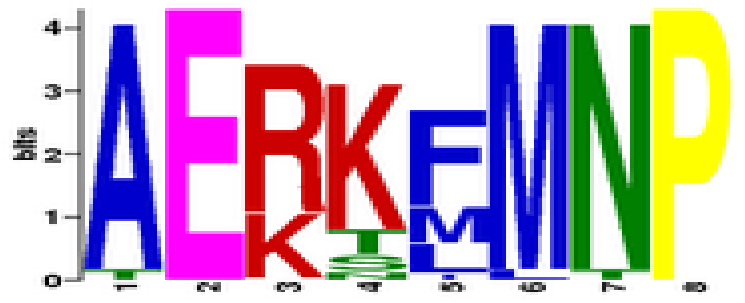

Regex:
AE[RK]K[FM]MNP

## Motif 19

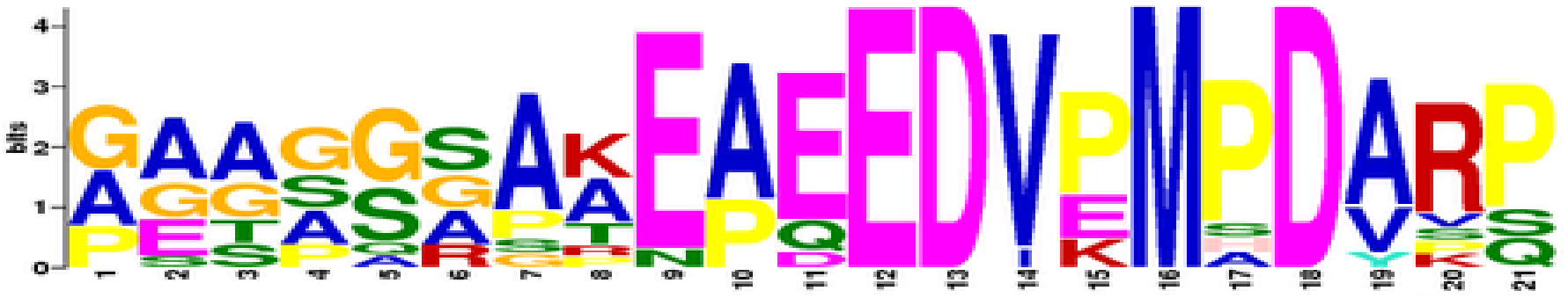

Regex:
[GAP][AEG][AG][GAS][GS][SAG]A[AK]E[AP]EEDV[PE]MPD[AV]RP

## Motif 20

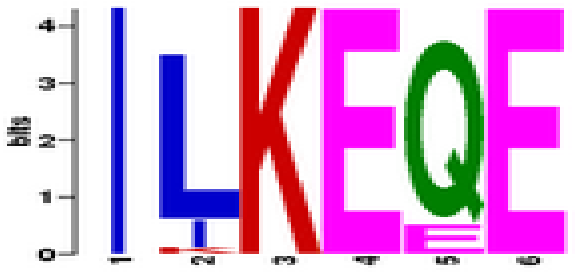

Regex:
ILKEQE

## Motif 21

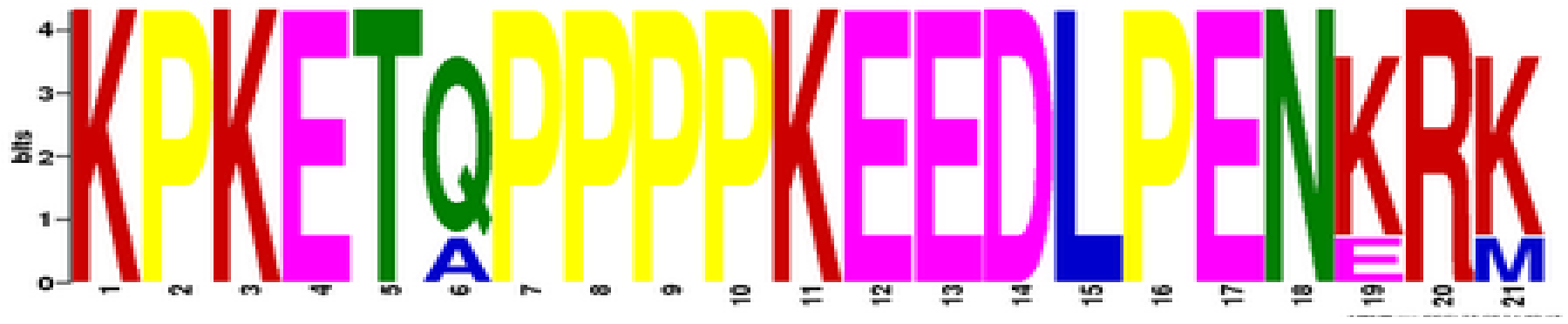

Regex:
KPKET[QA]PPPPKEEDLPEN[KE]R[KM]

**Motif 22**

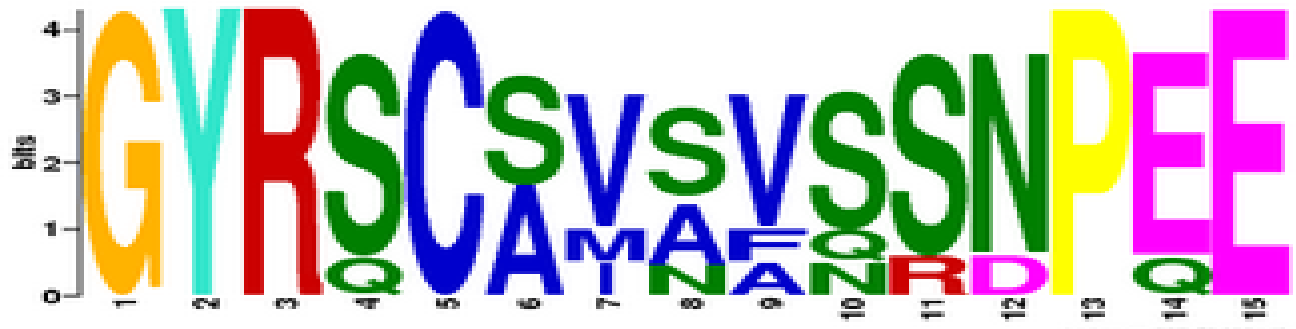

Regex:
GYRSC[AS]V[SA]VSSNPEE

**Motif 23**

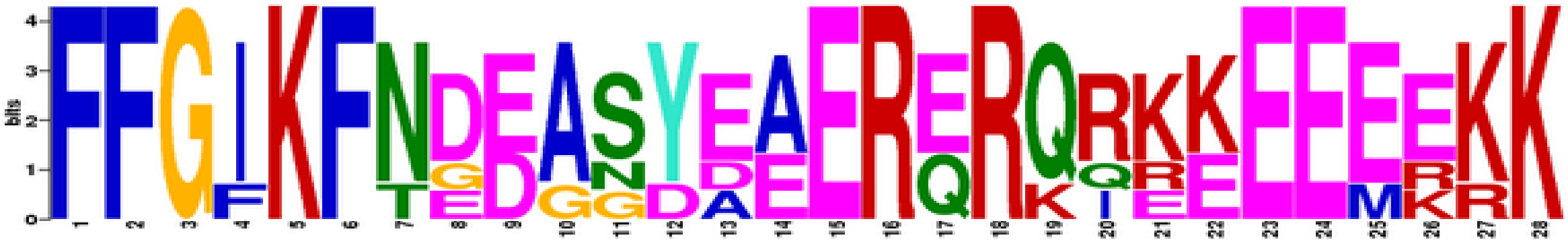

Regex:
FFG[IF]KF[NT][DEG][ED][AG][SGN][YD][EAD][AE]ER[EQ]R[QK][RIQ][KER][KE]EE[EM][EKR][KR]K

**Motif 24**

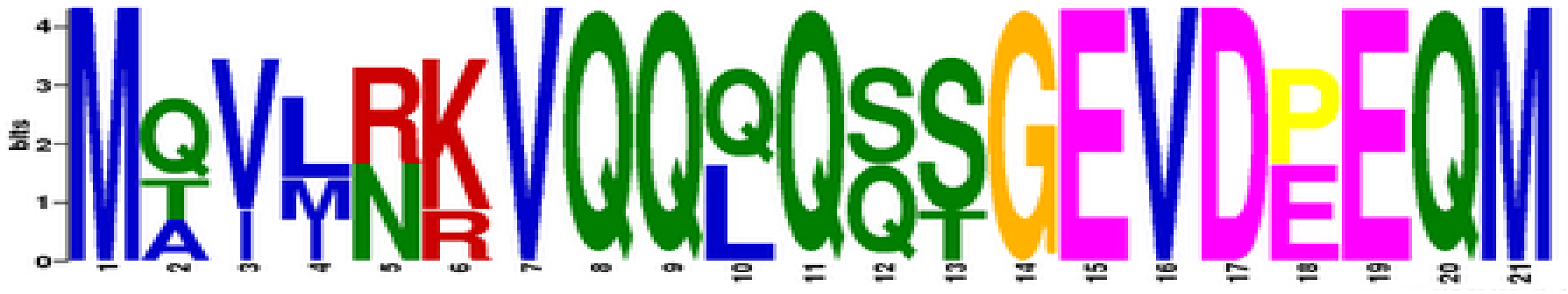

Regex:
M[QAT][VI][LIM][NR][KR]VQQ[LQ]Q[QS][ST]GEVD[EP]EQM

**Motif 25**

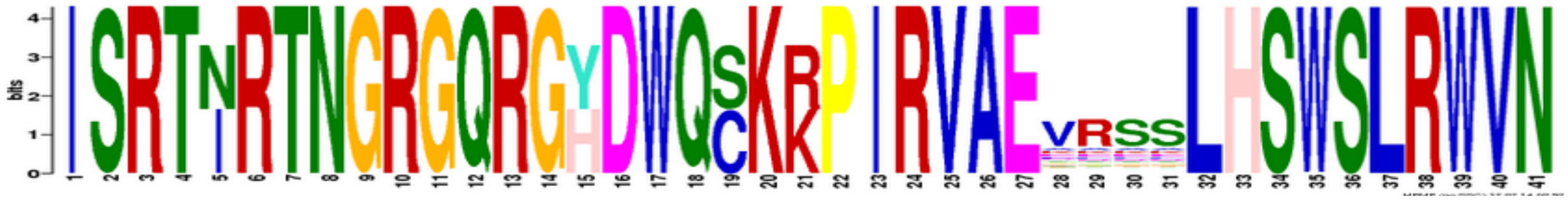

Regex:
ISRT[IN]RTNGRGQRG[HY]DWQ[CS]K[KR]PIRVAEVRSSLHSWSLRWVN

## Motif 26

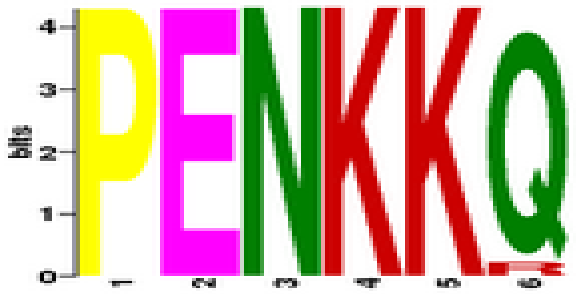

Regex:
PENKKQ

---

## Motif 27

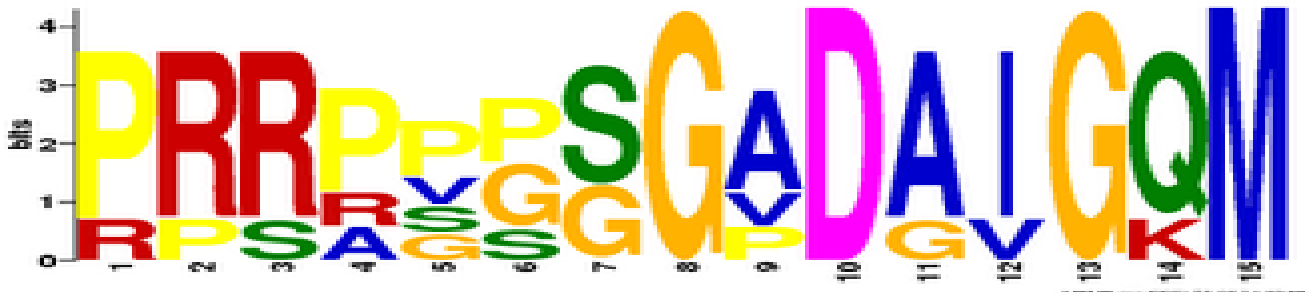

Regex:
[PR][RP][RS][PAR][PGSV][GPS][SG]G[APV]D[AG][IV]G[QK]M

---

## Motif 28

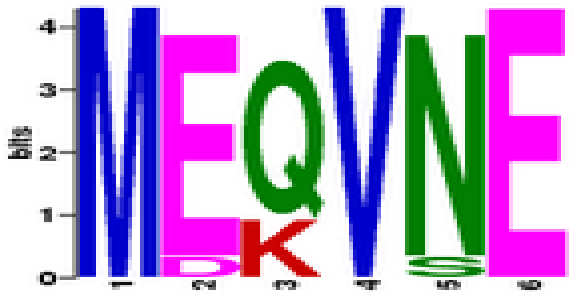

Regex:
ME[QK]VNE

---

## Motif 29

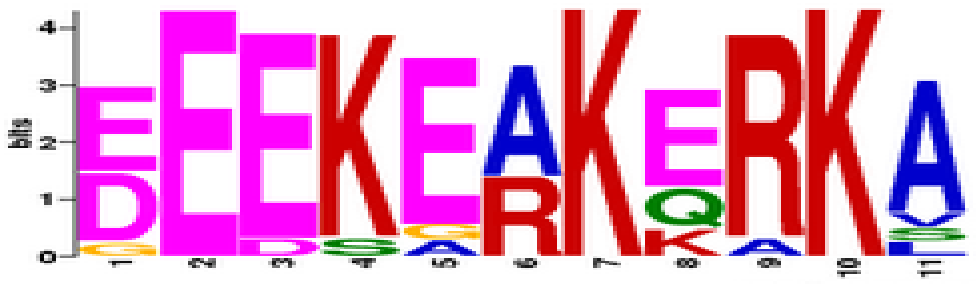

Regex:
[ED]EEKE[AR]K[EQ]RKA

---

**Motif 30**

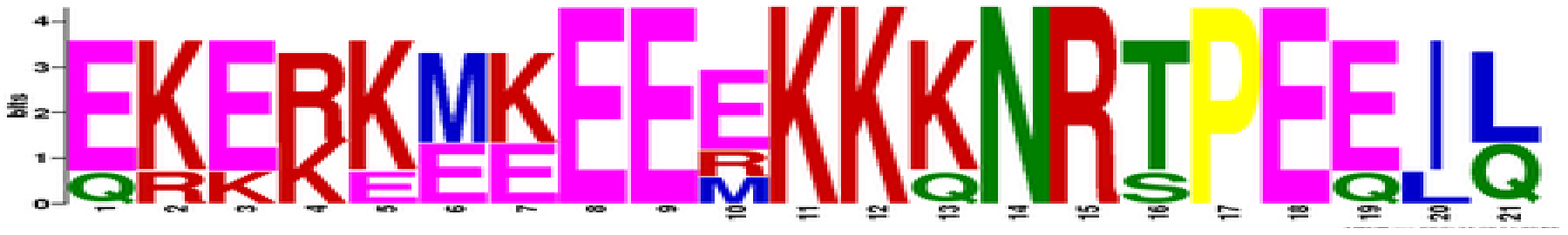

Regex:
[EQ][KR][EK][RK][KE][ME][KE]EE[EMR]KK[KQ]NR[TS]PE[EQ][IL][LQ]

**Supplementary Data 2C:**

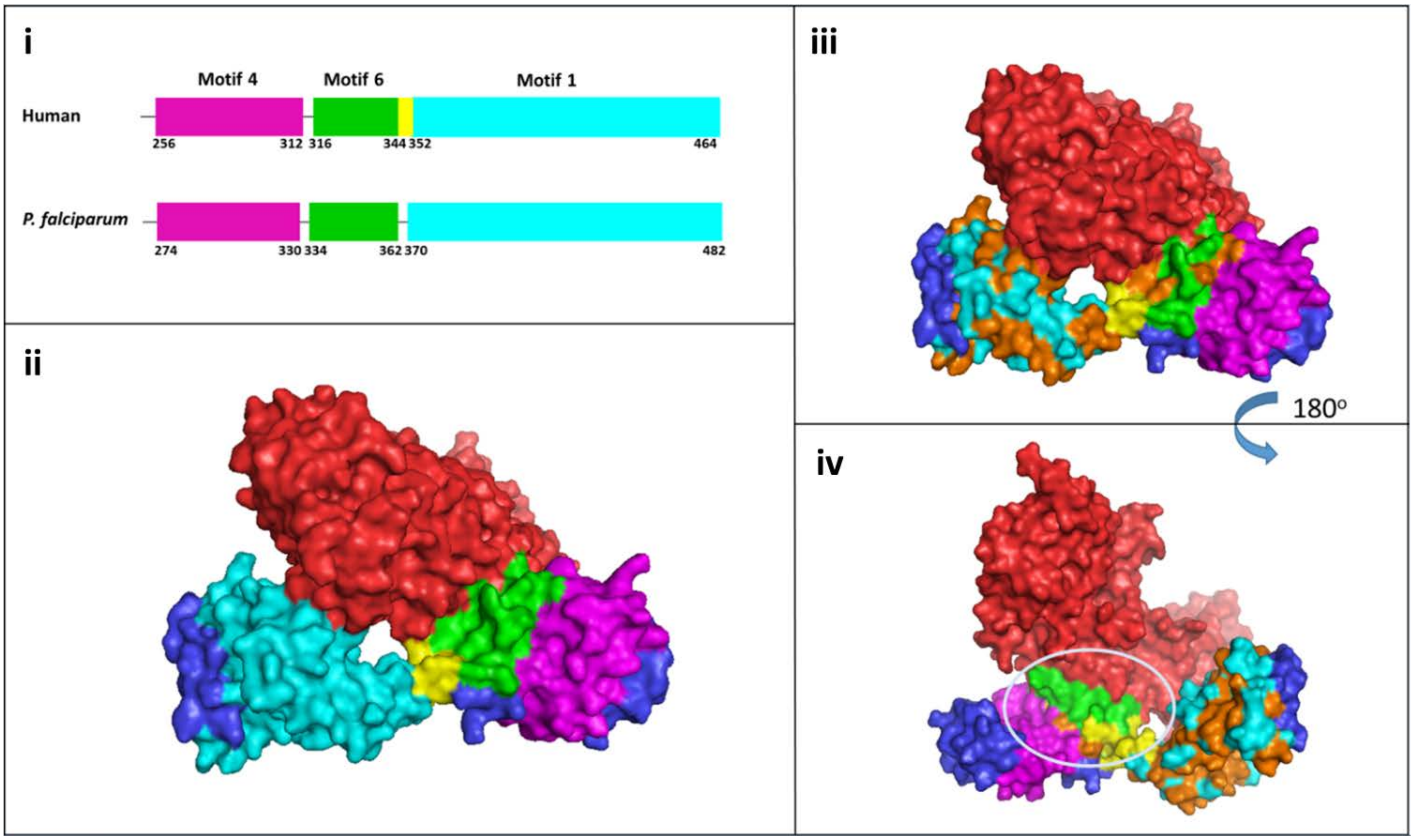

## Supplementary Data 3A:

| Convex Analysis | | | |
|---|---|---|---|
| model/complex name | DOPE Z-score | MetaMQAPII (Predicted GDT_TS) | HADDOCK score for cluster (if available) |
| | | | |
| DOCKED Complexes | | | |
| Alpha_redocked_CYS.pdb | -0.82 | 70.50 | -129.9 +/- 7.4 |
| Alpha_redocked_SER.pdb | -0.88 | 66.94 | -127.5 +/- 5.6 |
| Beta_redocked_CYS.pdb | -0.86 | 71.33 | -101.3 +/- 1.4 |
| Beta_redocked_SER.pdb | -0.89 | 71.21 | -158.8 +/- 7.8 |
| Pf_redocked_CYS.pdb | -0.80 | 66.38 | -93.7 +/- 2.5 |
| Pf_redocked_SER.pdb | -0.79 | 67.81 | -128.3 +/- 4.9 |
| template_resdocked_Schmid_SER.pdb | -1.01 | 67.80 | -119.2 +/- 3.5 |
| template_redocked_Schmid_CYS.pdb | -1.07 | 73.49 | -124.3 +/- 2.1 |
| | | | |
| MODELED COMPLEXES | | | |
| Alpha_modeled_CYS.pdb | -0.55 | 69.63 | |
| Alpha_modeled_SER.pdb | -0.55 | 69.52 | |
| Beta_modeled_CYS.pdb | -0.58 | 69.82 | |
| Beta_modeled_SER.pdb | -0.60 | 68.91 | |
| Pf_modeled_CYS.pdb | -0.44 | 65.20 | |
| Pf_modeled_SER.pdb | -0.47 | 68.61 | |
| TEMPLATES | | | |
| Schmid_CYS_template.pdb | -1.04 | 73.52 | |
| Schmid_SER_template.pdb | -1.06 | 73.30 | |

**Supplementary Data 3B:**

| Concave Analysis | | |
|---|---|---|
| model/complex name | DOPE Z-score | MetaMQAPII (Predicted GDT_TS) |
| | | |
| **HopTPR1-Hsp70$_{GPTIEEVD}$** | | |
| PfTPR1_PfHsp70.pdb<br>HsTPR1_HsHsp70.pdb<br>1ELW_trunc.pdb (template) | -2.24<br>-2.32<br>-2.57 | 72.20<br>81.45<br>78.40 |
| **HopTPR2AB-Hsp70$_{EVD}$-Hsp90$_{MEEVD}$** | | |
| PfTPR2AB_PfHsp70.pdb<br>HsTPR2AB_HsHsp70_HsHsp90.pdb<br>3UQ3.pdb (template) | -0.94<br>-1.01<br>-1.52 | 47.10<br>53.46<br>56.77 |
| **HopTPR2B-Hsp70$_{PTVEEVD}$** | | |
| PfTPR2B_PfHsp70$_{PTVEEVD}$<br>HsTPR2B_HsHsp70$_{PTVEEVD}$<br>3UPV.pdb (template) | -1.83<br>-1.55<br>-2.35 | 54.17<br>63.17<br>81.25 |

**Supplementary Data 4A:**

| Hsp90 models | | | | |
|---|---|---|---|---|
| **modelled** | **PfHsp90** | **HsHsp90-alpha** | **HsHsp90-beta** | **ScHsp90** |
| | | | | |
| 1 | Q312 | T293 | T285 | T273 |
| 2 | K313 | K294 | K286 | K274 |
| 3 | P314 | P295 | P287 | P275 |
| 15 | E326 | E307 | E299 | E287 |
| 16 | E327 | E308 | E300 | E288 |
| 19 | S330 | E311 | E303 | A291 |
| 20 | F331 | F312 | F304 | F292 |
| 22 | K333 | K314 | K306 | K294 |
| 26 | N337 | N318 | N310 | N298 |
| 27 | D338 | D319 | D311 | D299 |
| 28 | W339 | W320 | W312 | W300 |
| 29 | E340 | E321 | E313 | E301 |
| 30 | D341 | D322 | D314 | D302 |
| 53 | K364 | R345 | R337 | K325 |
| 70 | K381 | K362 | K354 | K342 |
| 72 | Y383 | Y364 | Y356 | Y344 |
| 76 | V387 | V368 | V360 | V348 |
| 77 | F388 | F369 | F361 | F349 |
| 79 | M390 | M371 | M363 | T351 |
| 139 | K450 | K431 | K423 | K411 |
| 140 | E451 | E432 | E424 | E412 |
| 143 | K454 | K435 | K427 | E415 |
| 144 | K455 | K436 | K428 | K416 |
| 147 | E458 | E439 | E431 | S419 |
| 176 | S487 | S468 | S460 | T448 |
| 177 | K488 | A469 | Q461 | K449 |
| 178 | S489 | S470 | S462 | S450 |
| 180 | D491 | D472 | D464 | D452 |
| 181 | E492 | E473 | E465 | E453 |
| 187 | E498 | D479 | E471 | D459 |
| 190 | D501 | T482 | S474 | T462 |
| 191 | R502 | R483 | R475 | R463 |
| 193 | K504 | K485 | K477 | P465 |
| 194 | E505 | E486 | E478 | E466 |

| Hop models | | | |
|---|---|---|---|
| modelled | PfHop | HsHop | ScHop |
| | | | |
| 73 | I315 | R297 | K333 |
| 77 | K319 | K301 | K337 |
| 100 | R342 | N324 | Q360 |
| 103 | L345 | L327 | L363 |
| 105 | E347 | E329 | E365 |
| 106 | D348 | H330 | H366 |
| 107 | N349 | R331 | R367 |
| 108 | N350 | T332 | T368 |
| 109 | R351 | P333 | A369 |
| 110 | A352 | D334 | D370 |
| 112 | R354 | L336 | L372 |
| 113 | N355 | K337 | T373 |
| 116 | K358 | Q340 | R376 |
| 119 | E361 | E343 | E379 |
| 120 | R362 | K344 | K380 |
| 123 | E365 | K347 | K383 |
| 127 | K369 | R351 | A387 |
| 151 | D393 | D375 | D411 |
| 152 | F394 | Y376 | W412 |
| 153 | P395 | P377 | P413 |
| 154 | N396 | Q378 | N414 |
| 156 | K398 | M380 | V416 |
| 157 | K399 | K381 | K417 |
| 160 | D402 | T384 | T420 |
| 164 | R406 | K388 | K424 |
| 176 | R418 | R400 | R436 |
| 180 | L422 | Y404 | L440 |
| 183 | L425 | L407 | L443 |
| 184 | I426 | L408 | M444 |
| 185 | E427 | E409 | S445 |
| 186 | Y428 | F410 | F446 |
| 187 | P429 | Q411 | P447 |
| 188 | S430 | L412 | E448 |
| 191 | E433 | K415 | A451 |
| 192 | D434 | D416 | D452 |
| 217 | M459 | M441 | V477 |

**Supplementary Data 4B:**

| Human complexes vs yeast template complex | | | | | |
|---|---|---|---|---|---|
| **Conserved** | **alpha and beta** | **template and beta** | **template only** | **alpha only** | **beta only** |
| 15[I/H]157 | 28[P]188 | 180[H]120 | 139[H]119 | 178[H]116 | 177[H]113 |
| 26[H]183 | 139[I]119 | | 177[H]109 | | |
| 26[H]185 | 143[I]119 | | 180[I/H]116 | | |
| 28[P]156 | | | 187[I]116 | | |
| 28[I]176 | | | 193[P]109 | | |
| 28[P]180 | | | 194[I]110 | | |
| 30[I/H]157 | | | | | |
| 140[I/H]123 | | | | | |
| 180[I]120 | | | | | |
| 181[I/H]120 | | | | | |
| 190[H]108 | | | | | Non-specific |
| 194[I/H]77 | | | | | Hsp90: 187 |
| 194[I]106 | | | | | Hop: 107, 116 |
| | | | | | |
| Total: 18 | Total: 3 | Total: 1 | Total: 7 | Total: 1 | Total: 4 |

| Human complexes vs *P. falciparum* complexes | | | | | |
|---|---|---|---|---|---|
| **conserved** | **alpha and beta** | **Pf and beta** | **Pf only** | **alpha only** | **beta only** |
| 15[I/H]157 | 26[H]183 | 177[H]113 | 143[I]123 | 178[H]116 | 180[H]120 |
| 28[I]176 | 26[H]185 | | 177[H]109 | | |
| 28[P]180 | 28[P]156 | | 180[I/H]116 | | |
| 30[I/H]157 | 28[P]188 | | 187[I/H]112 | | |
| 143[I]119 | 139[I]119 | | 190[I]112 | | |
| 180[I]120 | 140[I/H]123 | | 194[H]108 | | |
| 181[I]120 | 181[H]120 | | | | |
| 190[H]108 | 194[I]106 | Non-specific | Non-specific | | Non-specific |
| 194[I/H]77 | | | Hsp90: 26, 140 | | Hsp90: 187 |
| | | Hop: 107 | Hop: 156 | | Hop: 116 |
| | | | | | |
| Total: 12 | Total: 9 | Total: 2 | Total: 11 | Total: 1 | Total: 3 |

**Supplementary Data 4C:**

| Hsp90 Residues | | | |
|---|---|---|---|
| **template** | **alpha** | **beta** | **Pf** |
| 22 (K249) | | | |
| 28 (W300) | 28 (W320) | 28 (W312) | |
| 178 (S450) | 178 (S470) | | 178 (S489) |
| 180 (D452) | | | |
| 191 (R463) | 191 (R483) | 191 (R475) | 191 (R502) |
| 194 (E466) | | | 194 (E505) |
| **Hop Residues** | | | |
| | | | 108 (N350) |
| | | | 112 (R354) |
| 116 (R376) | 116 (Q340) | | |
| | 160 (T284) | 160 (T284) | |
| 176 (R436) | 176 (R400) | 176 (R400) | 176 (R418) |

# Supplementary Data 5A:

| TPR1-Hsp70$_{GPTIEEVD}$ | | | | |
|---|---|---|---|---|
| **Human and Pf** | | **Human only** | | **Pf only** |
| K5[H/I]D(-1) | | K5[H/I]E(-3) | | C77[H]T(-6) |
| N9[H]D(-1) | | K5[I]E(-4) | | S104[H]E(-4) |
| L12[P]V(-2) | | A43[P]V(-2) | | |
| Y24[P]V(-2) | | A43[P]V(-5) | | |
| N40[H]D(-1) | | K47[I]E(-3) | | |
| A46[P]V(-5) | | A77[P]P(-7) | | |
| K70[H/I]D(-1) | | F81[P]P(-7) | | |
| K70[H]E(-3) | | | | |
| K70[I]E(-4) | | | | |
| R74[I]D(-1) | | | | |
| R74[H]E(-3) | | | | |
| R74[H/I]E(-4) | | | | |
| | | | | |
| **TPR2A-Hsp90$_{MEEVD}$** | | | | |
| **Human and Pf** | | **Human only** | | **Pf only** |
| N9[H]D(-1) | | K5[I]D(-1) | | K15[H/I]D(-1) |
| Y12[H]E(-4) | | T36[H]D(-1) | | K15[I]E(-3) |
| Y12[P]V(-2) | | K77[H]E(-3) | | K15[H]M(-5) |
| Y12[H]M(-5) | | N84[H]E(-4) | | R81[I]D(-1) |
| N40[H]D(-1) | | | | |
| A43[P]V(-2) | | | | |
| E47[H]E(-4) | | | | |
| K77[H/I]D(-1) | | | | |
| K77[I]E(-3) | | | | |
| R81[H]E(-3) | | | | |
| R81[H/I]E(-4) | | | | |
| | | | | |
| **TPR2B-Hsp70$_{EVD}$** | | | | |
| **Human and Pf** | | **Human only** | | **Pf only** |
| F12[P]V(-2) | | Q13[H]E(-3) | | |
| Y24[P]V(-2) | | | | |
| N40[H]D(-1) | | | | |
| A43[P]V(-2) | | | | |
| R74[I]D(-1) | | | | |
| R74[H]E(-3) | | | | |
| | | | | |
| **TPR2B-Hsp70$_{PTVEEVD}$** | | | | |
| **Human and Pf** | | **Human only** | | **Pf only** |
| F12[P]V(-2) | | K70[H]D(-1) | | S73[H]T(-6) |
| Y24[P]V(-2) | | R74[H]T(-6) | | |
| N40[H]D(-1) | | A77[P]V(-5) | | |
| A43[P]V(-2) | | | | |
| K47[H/I]E(-4) | | | | |
| K70[I]D(-1) | | | | |
| K70[I]E(-3) | | | | |
| R74[I]D(-1) | | | | |
| R74[H]E(-3) | | | | |
| Q110[H]T(-6) | | | | |

**Supplementary Data 5B:**

| **TPR1-Hsp70$_{GPTIEEVD}$** | | **TPR2A-Hsp90$_{MEEVD}$** | | **TPR2B-Hsp70$_{EVD}$** | | **TPR2B-Hsp70$_{PTVEEVD}$** |
|---|---|---|---|---|---|---|
| K5[H/I]D(-1) | | | | | | |
| N9[H]D(-1) | | N9[H]D(-1) | | | | |
| L12[P]V(-2) | | Y12[H]E(-4) | | F12[P]V(-2) | | F12[P]V(-2) |
| | | Y12[P]V(-2) | | | | |
| | | Y12[H]M(-5) | | | | |
| Y24[P]V(-2) | | | | Y24[P]V(-2) | | Y24[P]V(-2) |
| N40[H]D(-1) | | N40[H]D(-1) | | N40[H]D(-1) | | N40[H]D(-1) |
| | | A43[P]V(-2) | | A43[P]V(-2) | | A43[P]V(-2) |
| A46[P]V(-5) | | | | | | |
| | | E47[H]E(-4) | | | | K47[H/I]E(-4) |
| K70[H/I]D(-1) | | *K77[H/I]D(-1) | | | | K70[I]D(-1) |
| K70[H]E(-3) | | *K77[I]E(-3) | | | | K70[I]E(-3) |
| K70[I]E(-4) | | | | | | |
| R74[I]D(-1) | | *R81[H]E(-3) | | R74[I]D(-1) | | R74[I]D(-1) |
| R74[H]E(-3) | | *R81[H/I]E(-4) | | R74[H]E(-3) | | R74[H]E(-3) |
| R74[H/I]E(-4) | | | | | | |
| | | | | | | Q110[H]T(-6) |

**Supplementary Data 5C:**

```
                        10        20        30        40        50        60
                    ....|....|....|....|....|....|....|....|....|....|....|....|
PfTPR1      7   AQRLKELGNKCFQEGKYEEAVKYFSDAITNDPLDHVLYSNLSGAFASLGRFYEALESANKCISIK  71
HsTPR1      4   VNELKEKGNKALSVGNIDDALQCYSEAIKLDPHNHVLYSNRSAAYAKKGDYQKAYEDGCKTVDLK  67

PfTPR2A   243   GDEHKLKGNEFYKQKKFDEALKEYEEAIQINPNDIMYHYNKAAVHIEMKNYDKAVETCLYAIENR  307
HsTPR2A   225   ALKEKELGNDAYKKKDFDTALKHYDKAKELDPTNMTYITNQAAVYFEKGDYNKCRELCEKAIEVG  289
ScTPR2A   262   ADKEKAEGNKFYKARQFDEAIEHYNKAWELH-KDITYLNNRAAAEYEKGEYETAISTLNDAVEQG  325

PfTPR2B   378   AEEHKNKGNEYFKNNDFPNAKKEYDEAIRRNPNDAKLYSNRAAALTKLIEYPSALEDVMKAIELD  442
HsTPR2B   360   ALEEKNKGNECFQKGDYPQAMKHYTEAIKRNPKDAKLYSNRAACYTKLLEFQLALKDCEECIQLE  424
ScTPR2B   396   AEEARLEGKEYFTKSDWPNAVKAYTEMIKRAPEDARGYSNRAAALAKLMSFPEAIADCNKAIEKD  460

                      70        80        90       100       110       120
                ....|....|....|....|....|....|....|....|....|....|....|
PfTPR1     72   KDW-------PKGYIRKGCAEHGLRQLSNAEKTYLEGLKIDPNNKSLQDALS--KVRNE-  121
HsTPR1     68   PDW-------GKGYSRKAAALEFLNRFEEAKRTYEEGLKHEANNPQLKEGLQ--NMEARL  117

PfTPR2A   308   YNFKAEFIQVAKLYNRLAISYINMKKYDLAIEAYRKSLVEDNNRATRNALKELERRKE--  365
HsTPR2A   290   RENREDYRQIAKAYARIGNSYFKEEKYKDAIHFYNKSLAEHRTPDVLKKCQQAEKILK--  347
ScTPR2A   326   REMRADYKVISKSFARIGNAYHKLGDLKKTIEYYQKSLTEHRTADILTKLRNAEKELK--  383

PfTPR2B   443   PTF-------VKAYSRKGNLHFFMKDYYKALQAYNKGLELDPNNKECLEGYQ--RCAFKI  493
HsTPR2B   425   PTF-------IKGYTRKAAALEAMKDYTKAMDVYQKALDLDSSCKEAADGYQ--RCMMAQ  475
ScTPR2B   461   PNF-------VRAYIRKATAQIAVKEYASALETLDAARTKDAEVNNGSSAREIDQLYYKA  513
```